\documentclass[%
 reprint,
 amsmath,amssymb,
 aps,
]{revtex4-1}

\usepackage{graphicx}
\usepackage{dcolumn}
\usepackage{bm}
\usepackage{subfigure}

\usepackage{hhline}
\usepackage{multirow}

\begin{document}

\preprint{APS/123-QED}

\def\mean#1{\left< #1 \right>} 
\title{Tune compensation in nearly scaling Fixed Field alternating gradient Accelerators}

\author{M. Haj Tahar}%
 \email{hajtaham@gmail.com}
\affiliation{%
CERN, Geneva, Switzerland 
}%

\author{F. M\'eot}
\affiliation{Collider Accelerator Department, Brookhaven National Laboratory, Long Island, Upton, NY, USA
}%

\begin{abstract}
In this paper, we investigate the stability of the particle trajectories in Fixed Field alternating gradient Accelerators (FFA) in the presence of field errors: The emphasis is on the scaling radial sector FFA type: a collaboration work is on-going in view of better understanding the properties of the 150 MeV scaling FFA at KURNS in Japan,  and progress towards high intensity operation. Analysis of certain types of field imperfections revealed some interesting features about this machine that explain some of the experimental results and generalize the concept of a scaling FFA to a non-scaling one for which the tune variations obey a well defined law. A compensation scheme of tune variations in imperfect scaling FFAs is presented. This is the cornerstone of a novel concept of a non-linear non-scaling radial sector fixed tune FFA that we present and discuss in details in the last part of this paper.
\end{abstract}

\pacs{Valid PACS appear here}
\maketitle


\section{Introduction}
Scaling Fixed Field alternating gradient Accelerator (FFA) is a concept that was invented in the 1950s almost independently in the US, Japan and USSR \cite{symon,ohkawa,kolomensky}. Several electron machines were built in the US. However, it was not until the 1990s that the interest for scaling FFAs was revived in Japan. Several machines were built among which a \mbox{150 MeV} proton machine at Kyoto University Institute for Integrated Radiation and Nuclear Science (KURNS) \cite{note}. One main feature of this machine is its potential for high power applications, hence its use as a proton driver for an Accelerator Driven Sub-critical Reactor (ADSR) \cite{first_injection} at KURNS. A large dynamic acceptance can be achieved since the crossing of the betatron resonances is avoided in this concept. This is achieved by introducing $R^k$ increase of the magnetic field with the radius resulting in the beam experiencing the same focusing throughout the acceleration, therefore keeping the tunes constant. The magnetic field profile allowing this writes in cylindrical coordinates in the form  $B = B_0 (R/R_0)^k F(\theta)$ where $B$ is the vertical component of the magnetic field in the median plane, $R$ is the radial coordinate with respect to the center of the ring, $B_0$ the reference field at $R=R_0$, $F(\theta)$ the flutter function describing the azimuthal variations of the field and $k$ is the average field index of the magnets, sometimes also referred to as the scaling factor, defined by $k=R/B . \partial B/ \partial R$ and ideally a constant value everywhere in the ring. One shall insist here that this form of the magnetic field is a sufficient but non-necessary condition in order to obtain a fixed tune FFA. In the scaling FFA concept, the orbit excursion is uniquely determined by the field index and the flutter function. Given that all built machines in the past opted for a phase advance per cell below $180^\circ$, the orbit shift is therefore large, of the order of $1$ meter which makes the magnet larger than typical synchrotron magnets. Besides, due to the complexity of the field profile and flutter of the magnets, field imperfections can be problematic and difficult to cure since the orbits move outwards from the centre of the machine with increasing energies. This can lead to the crossing of several betatron resonances at low speed and to beam deterioration in consequence. For instance, this is a strong contribution to the overall low beam transmission in the KURNS FFA \cite{Ishi_FFAG,suzie_charac}.

Despite some controversy regarding the definition of FFA, it is worth mentioning that spiral sector cyclotrons are a class of non scaling FFA where the alternating gradient effect comes from the spiral shaped poles \cite{symon}. Some of these machines existed since the 50s such as the PSI and TRIUMF cyclotrons which have been operating for over 40 years at high intensity regime with very low losses. In this concept, the scaling is sacrificed in favor of isochronism thus allowing such machines to operate in continuous wave mode. Unlike cyclotrons, scaling FFAs operate like synchro-cyclotrons in that the rf frequency must be varied to remain synchronous with the accelerating particles.\\
In addition, although FFA are generally assumed to have non linear field profile with the radius, there exists a second class of FFAs, that of linear concepts. The linear FFAs are so called because they only use linear elements such as dipoles and quadrupoles. A proof-of-principle electron model of this concept called EMMA \cite{EMMA} has been constructed and operated successfully at the STFC Daresbury Laboratory in the UK. Such a concept has been proposed in the context of rapid acceleration of unstable muons for future high energy colliders. Furthermore, the first Energy Recovery Linac (ERL) based on linear FFA magnets is being constructed at Cornell University \cite{cbeta} as a prototype for a potential Electron Ion collider, so called eRHIC \cite{erhic}.

\begin{figure}
\centering 
\includegraphics*[width=8cm]{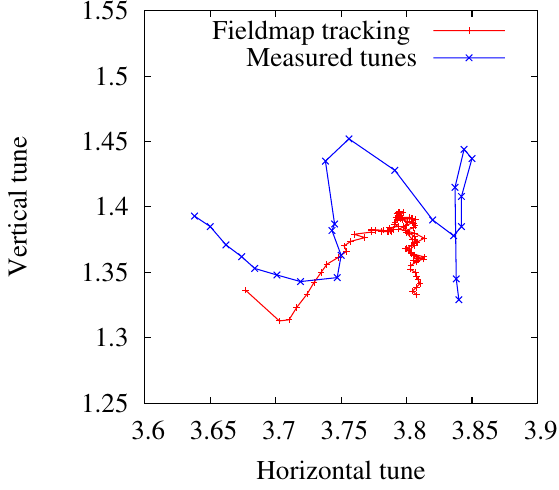}
\caption{Betatron tunes from 11 to 100 MeV (left to right). The details of the measurement as well as the methods and tools
that were used to characterize the 150 MeV KURNS FFA are discussed in \cite{suzie_charac}.}
\label{fig:tunes_meas}
\end{figure}

In the original work of Symon \cite{symon}, it was established that if the average field index is kept constant and the closed orbits are geometrically similar in a way that will become clearer later on in this paper, then the number of betatron oscillations can be approximated by: 
\begin{eqnarray}
\begin{cases}
& \nu_x ^2 \approx k+1   \\  \\
& \nu_y ^2 \approx -k+\mathcal{F}^2 \left(1 + 2 \tan^2(\xi) \right)
\label{eq:symon_tunes}
\end{cases}
\end{eqnarray} 
where $\mathcal{F}$ is the magnetic flutter and $\xi$ is the spiral angle of the magnets. Nevertheless, due to field imperfections, it is particularly challenging to design and manufacture a magnet which produces exactly the desired magnetic field. For instance, the measured as well as the simulated tunes of the 150 MeV FFA at KURNS are shown in \mbox{fig. \ref{fig:tunes_meas}} where one can observe non negligible tune variations.

In what folllows, we will derive the linearized Hills equations of motion in cylindrical coordinates in order to take into account the field imperfections and define the domain of validity of the above approximation. One objective of this paper is to establish general expressions of the ring tunes that take into account the non-scaling of the orbits due to field imperfections and compare with the measured as well as the simulated values. This will be a crucial result in order to establish a correction scheme to minimize the tune variations in imperfect radial scaling FFA.

\section{Geometry of the closed orbit}

\begin{figure}
\centering 
\includegraphics*[width=8cm]{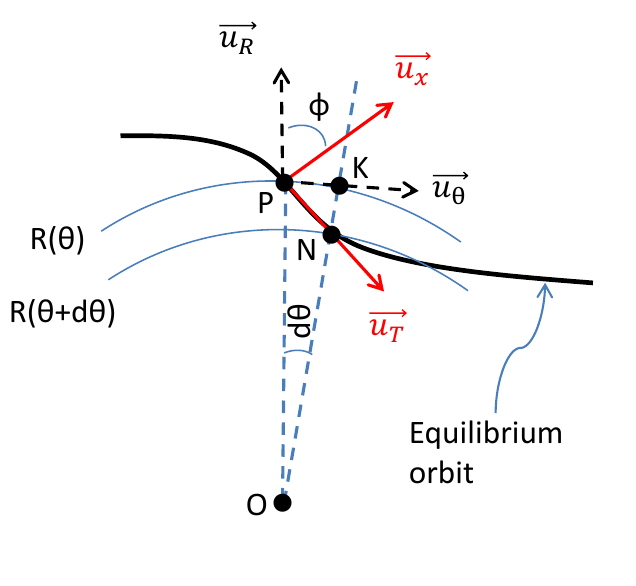}
\caption{Geometric properties of the closed orbit. The point O is the center of the ring and the particle motion is clockwise.}
\label{fig:closed_orbit}
\end{figure}

In cylindrical coordinates as shown in fig. \ref{fig:closed_orbit}, it can be shown that any median plane closed orbit satisfies the following equation:
\begin{eqnarray}
\tan(\phi)=-\dfrac{dR / d\theta}{R} = -\dfrac{\dot{R}}{R} \label{Eq:tan}
\end{eqnarray}
where $\phi(\theta)$ is the angle between the extended radial line and the normal to the equilibrium orbit, \mbox{i.e. $\phi=(\widehat{\overrightarrow{u_R},\overrightarrow{u_x}})$.} \\ $\phi$ is a natural parameter to describe the scalloping of the closed orbit (note that $\phi=-T$ for the element ``polarmes'' in the tracking code zgoubi \cite{zgoubi}): if $\phi(\theta)=0$ for all $\theta$, then the closed orbit is a circle. In order for $R$ to be a well defined function of the azimuthal angle, the orbit cannot be moving radially outwards. This sets the following condition on $\phi$: $|\phi| < \pi/2$.
 \\
Differentiating again with respect to the azimuthal angle, one obtains:
\begin{eqnarray}
\ddot{R}=-\dot{R} \tan(\phi)-R\dot{\phi} \left[1+\tan^2(\phi)\right]
\end{eqnarray}

Now, one can express the signed curvature of the closed orbit as  a function of the radius:
\begin{eqnarray}
\dfrac{1}{\rho} = \dfrac{R^2+2\left(\dot{R}\right)^2-R \ddot{R}}{\left[R^2+\left(\dot{R}\right)^2\right]^{3/2}} = \dfrac{1}{R} \left(1+\dot{\phi}\right) \cos(\phi)
\label{Eq:rho}
\end{eqnarray}
where one uses the convention that the particle motion is clockwise with increasing $\theta$ and that the curvature is positive if the momentum vector turns clockwise with increasing $\theta$. \\
Thus, when defining the particle closed orbit in cylindrical coordinates, the particle radius $R$ changes with the azimuthal angle. Such a dependence is described by the angle $\phi(\theta)$ which is also defined as the angle between the tangent to the circle of radius $R(\theta)$ and the particle momentum vector at the azimuthal angle $\theta$. It results that the position of the particle on the closed orbit is specified by the variables $(R(\theta), \theta, z(\theta))$.  \\
Note that in a drift space where the particle trajectory is a straight line, Eq. (\ref{Eq:rho}) holds and one can easily show \mbox{that $\dot{\phi}=-1$} by observing that the vector $\overrightarrow{u_x}$ remains unchanged along a straight line. In addition, $(1+\dot{\phi})>0$ for positive curvature, which implies that the vertical component of the magnetic field points upwards for positive charged particles.
 
\subsection{Arclength in cylindrical and generalized azimuthal coordinates}
To describe the transverse beam dynamics in particle accelerators, very often the first approach consists in searching for the closed orbit corresponding to a specific energy and writing the linearized Hill's equation of motion describing the betatron oscillations around that orbit.   
In the seminal MURA paper of 1956, Symon introduced the concept of generalized azimuthal coordinates in which each orbit is specified by its equivalent radius $\mathcal{R}$ defined in the following way:
\begin{eqnarray}
\mathcal{C} = 2 \pi \mathcal{R}
\end{eqnarray}
where $\mathcal{C}$ is the closed orbit length for a specific energy. In addition, a generalized azimuthal coordinate is defined which is related to the distance measured along the orbit, i.e. to the curvilinear abscissa $s$ in the following way:
\begin{eqnarray}
s = \mathcal{R} \vartheta
\end{eqnarray}
In the formalism to be developed in the following section, one will write the linearized equations of motion in cylindrical coordinates which is particularly useful when calculating a correction scheme of magnetic field imperfections. For this reason, we will express the increment $ds$ in $s$ as a function of the increment $d\theta$ in $\theta$ along a closed orbit. \\
The increment of the arclength in cylindrical coordinates is given by:
\begin{eqnarray}
ds &=& \left(dR^2 + R^2 d\theta^2 + dz^2\right)^{1/2} \nonumber \\
&=& R \left[1+ \left(\dfrac{\dot{R}}{R}\right)^2 + \left(\dfrac{\dot{z}}{R}\right)^2 \right]^{1/2} d\theta \label{Eq:ds_incr}
\end{eqnarray}
Neglecting the changes in the vertical direction (the ideal orbit usually lies in the median plane of the accelerator), then one obtains by making use of Eq. (\ref{Eq:tan}):
\begin{eqnarray}
\dfrac{ds}{d\theta} &=& \dfrac{R(\theta)}{\cos(\phi)} \label{Eq:dsdtheta1}\\
\dfrac{d^2s}{d\theta ^2} &=& \dfrac{\dot{R}}{\cos(\phi)} + \dfrac{R\dot{\phi}\tan(\phi)}{\cos(\phi)} \label{Eq:transform}
\end{eqnarray}
Note that Eq. (\ref{Eq:dsdtheta1}) can be also established by applying the sinus law in the triangle OPN in fig. \ref{fig:closed_orbit} and taking the limit $d\theta \rightarrow 0$.
Equating the change of the arclength in both coordinates yields:
\begin{eqnarray}
ds = \mathcal{R} d\vartheta = \dfrac{R}{\cos(\phi)} d\theta
\label{Eq:dsdtheta}
\end{eqnarray} 
so that one can establish the transformation from the generalized azimuthal coordinate to the cylindrical azimuthal coordinate:
\begin{eqnarray}
d\vartheta = \dfrac{R}{\mathcal{R}} \dfrac{1}{\cos(\phi)} d\theta \label{Eq:transf}
\end{eqnarray}
and 
\begin{eqnarray}
\vartheta = \dfrac{1}{\mathcal{R}} \int_0^ \theta \dfrac{R(u)}{\cos(\phi)} du
\end{eqnarray}
where one assumed that both quantities coincide \mbox{at $\theta=0$.} In particular, this shows that, even for a radial sector FFA, the generalized azimuthal angle $\vartheta$ and the polar angle $\theta$ are not identical. \\
In addition $\mathcal{R}$ can be defined in cylindrical coordinates as follows:
\begin{eqnarray}
\mathcal{R} = \dfrac{\mathcal{C}}{2\pi} = \dfrac{1}{2\pi} \int_0^{2\pi} \dfrac{R}{\cos(\phi)} d\theta
\end{eqnarray}
which shows that $\mathcal{R}$ will be larger than the mean radius $\mean{R}$ of the closed orbit. \\
Now, by making use of Eqs. (\ref{Eq:rho}) and (\ref{Eq:dsdtheta}), one can compute the average value of the curvature function for a given closed orbit in the following way:
\begin{eqnarray}
\dfrac{1}{2\pi} \int_0^{2\pi \mathcal{R}} \dfrac{ds}{\rho} &=&  \dfrac{1}{2\pi} \int_0^{2\pi} \dfrac{R(\theta)}{\rho(\theta)} \dfrac{1}{\cos(\phi)} d\theta \nonumber \\
&=& \dfrac{1}{2\pi} \int_0^{2\pi} \left(1+\dot{\phi}\right) d\theta = 1
\end{eqnarray}
as expected, since the sum of all deflecting angles of the closed orbit shall equal $2\pi$.

\section{Transverse equations of motion in cylindrical coordinates}
The transverse equations for the linear betatron oscillations around the closed orbit write in the following \mbox{way \cite{courant}:}
\begin{eqnarray}
 \dfrac{d^2x}{ds^2} + \dfrac{1-n}{\rho^2} x = 0
 \label{eq:eq_motion_frenet1}
\end{eqnarray}
\begin{eqnarray}
 \dfrac{d^2y}{ds^2} + \dfrac{n}{\rho^2} y = 0
\label{eq:eq_motion_frenet}
\end{eqnarray}
where $x$ and $y$ are the transverse deviations of the particle in the horizontal and vertical direction respectively. Since the field map in a fixed field accelerator is usually defined in cylindrical coordinates $(R,\theta)$ with respect to the center of the ring, it is natural to write the particle equations of motion in such a frame.  
Differentiating with respect to the azimuthal angle $\theta$, one obtains:
\begin{eqnarray}
\dfrac{dx}{d\theta} &=& \dfrac{dx}{ds} \dfrac{ds}{d\theta} \label{eq:dxdth} \\
\dfrac{d^2x}{d\theta^2} &=& \dfrac{d^2x}{ds^2} \left(\dfrac{ds}{d\theta}\right)^2 + \dfrac{dx}{ds}\dfrac{d^2s}{d\theta^2} \label{eq:d2xdth2} 
\end{eqnarray}
where the transverse coordinate is a function of $\theta$ only.
Now, injecting Eqs. (\ref{eq:dxdth}) and (\ref{eq:d2xdth2}) into the Hill's equations (\ref{eq:eq_motion_frenet1}) and (\ref{eq:eq_motion_frenet}) yields:
\begin{eqnarray}
 \ddot{x} -\dfrac{\ddot{s}}{\dot{s}} \dot{x} + \dfrac{\dot{s}^2}{\rho ^2}(1-n) x = 0
 \label{eq_motion_cylinx}
\end{eqnarray}
\begin{eqnarray}
 \ddot{y} -\dfrac{\ddot{s}}{\dot{s}} \dot{y} + \dfrac{\dot{s}^2}{\rho ^2}n y = 0
\label{eq_motion_cylin}
\end{eqnarray}
where the $\dot{()}$ represents the differentiation with respect to the azimuthal variable. The above equations can be written in the standard form:
\begin{eqnarray}
\ddot{u} + p(\theta) \dot{u} + q(\theta) u = 0  \label{stand}
\end{eqnarray}
This is a 2nd order linear differential equation with variable coefficients, for which, in general, a closed form solution is not known. The first order derivative is removed by making the Liouville-Green transformation:
\begin{eqnarray}
u(\theta)=v(\theta). \exp \left(-\dfrac{1}{2} \int_{\theta_0}^{\theta} p(h) dh \right)
\end{eqnarray}
which yields:
\begin{eqnarray}
\ddot{v}+\left(q(\theta)-\dfrac{1}{2}\dfrac{dp}{d\theta}-\dfrac{p^2}{4} \right) v=0  \label{d2v}
\end{eqnarray}
Now, back to Eqs. (\ref{eq_motion_cylinx}) and (\ref{eq_motion_cylin}). The above formalism applies by simply replacing $p(\theta)=-\ddot{s}/\dot{s}$ which yields:
\begin{eqnarray}
u(\theta) = \dfrac{\sqrt{ds/d\theta}}{\sqrt{ds/d\theta}_{\theta_0}} v(\theta) = C \sqrt{\dfrac{R(\theta)}{\cos(\phi)}} v(\theta)
\end{eqnarray}
and the differential equations for the linear betatron oscillations around the equilibrium orbit are given by:
\begin{eqnarray}
\begin{cases}
\ddot{v}_{x,y} + \left[ q_{x,y}(\theta) + f(\theta) \right] v_{x,y} = 0 \\  \\
q_x(\theta) = \dfrac{\dot{s}^2}{\rho ^2}(1-n) = \left(1+\dot{\phi}\right)^2 \left(1-n\right) \label{Eq:qx}\\ \\
q_y(\theta) = \dfrac{\dot{s}^2}{\rho ^2}n = \left(1+\dot{\phi}\right)^2 n \label{Eq:qy} \\ \\
p(\theta) = -\dfrac{\ddot{s}}{\dot{s}} = \left(1-\dot{\phi}\right) \tan(\phi) \\ \\
f(\theta) = -\dfrac{\dot{p}}{2} - \dfrac{p^2}{4}
\end{cases}
\end{eqnarray}
In the next section, one shall seek the expression of the field index in cylindrical coordinates.

\subsection{Field variations}
\subsubsection{Field index}
\begin{figure}
\centering 
\includegraphics*[width=8cm]{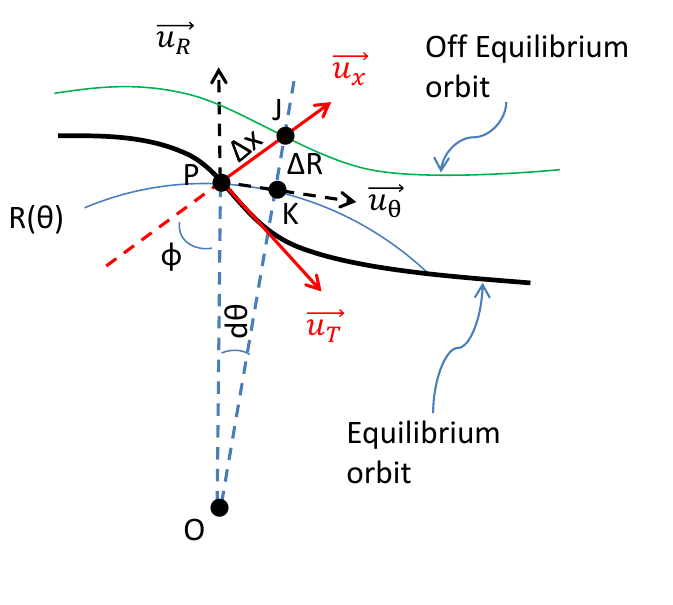}
\caption{Geometric properties of the particle motion around the closed orbit.}
\label{fig:drdx}
\end{figure}

Using the chain rule for partial differentiation, one obtains:
\begin{eqnarray}
n=-\dfrac{\rho}{B} \dfrac{\partial B}{\partial x} = -\dfrac{\rho}{B} \left[\dfrac{\partial B}{\partial R} \dfrac{\partial R}{\partial x} + \dfrac{\partial B}{\partial \theta} \dfrac{\partial \theta}{\partial x} \right] 
\label{Eq:field_index}
\end{eqnarray}
where $B$ is the magnetic field seen by the particle. \\
Applying the sinus law in the triangles PJK and POJ as illustrated in fig. \ref{fig:drdx}, one can establish \cite{cole}:
\begin{eqnarray}
\dfrac{\partial R}{\partial x} &=& \cos(\phi) \label{Eq:drdx} \\
\dfrac{\partial \theta}{\partial x} &=& \dfrac{\sin(\phi)}{R} \label{Eq:drdtheta}
\end{eqnarray}
Substituting Eqs. (\ref{Eq:rho}), (\ref{Eq:drdx}) and (\ref{Eq:drdtheta}) into Eq. (\ref{Eq:field_index}), one obtains:
\begin{eqnarray}
n = - \dfrac{R}{B} \dfrac{\partial B}{\partial R} \dfrac{1}{1+\dot{\phi}} - \dfrac{1}{B} \dfrac{\partial B}{\partial \theta} \dfrac{\tan(\phi)}{1+\dot{\phi}}
\end{eqnarray}
Now, the median plane magnetic field can be written in cylindrical coordinates in the general form:
\begin{eqnarray}
B(R,\theta) = B_m(R) F(R,\theta)
\end{eqnarray}
where $B_m(R)$ is the average magnetic field at a given radius which is R-dependent and $F(R,\theta)$ is the flutter function describing the azimuthal field variations along a fixed radius and satisfying 
\begin{eqnarray}
\dfrac{1}{2\pi} \int_0^{2\pi} F\left(R,\theta \right) d\theta = 1
\end{eqnarray} \\ \\
Taking the partial derivatives of the field in cylindrical coordinates yields:
\begin{eqnarray}
\dfrac{1}{B} \dfrac{\partial B}{\partial R} &=& \dfrac{1}{B_m(R)} \dfrac{\partial B_m}{\partial R} + \dfrac{1}{F(R,\theta)} \dfrac{\partial F(R,\theta)}{\partial R} \\
\dfrac{1}{B} \dfrac{\partial B}{\partial \theta} &=& \dfrac{1}{F(R,\theta)} \dfrac{\partial F(R,\theta)}{\partial \theta}
\end{eqnarray}
Finally, the field index can be expressed as follows:
\begin{eqnarray}
n = &-& \dfrac{R}{B_m} \dfrac{\partial B_m}{\partial R} \dfrac{1}{1+\dot{\phi}} - \dfrac{R}{F} \dfrac{\partial F}{\partial R} \dfrac{1}{1+\dot{\phi}} \nonumber \\
&-& \dfrac{1}{F} \dfrac{\partial F}{\partial \theta} \dfrac{\tan(\phi)}{1+\dot{\phi}}
\end{eqnarray}
and the character of the betatron oscillations is determined by Eqs. (\ref{Eq:qx}) which transform into:
\begin{widetext}
\begin{eqnarray}
q_x(\theta) &=& \left(1+\dot{\phi}\right)^2 + \left(1+\dot{\phi}\right) \dfrac{R}{B_m} \dfrac{\partial B_m}{\partial R} + \left(1+\dot{\phi}\right) \dfrac{R}{F} \dfrac{\partial F}{\partial R} + \left(1+\dot{\phi}\right) \tan(\phi) \dfrac{1}{F} \dfrac{\partial F}{\partial \theta} \label{Eq:qxtheta} \\
q_y(\theta) &=&  -\left(1+\dot{\phi}\right) \dfrac{R}{B_m} \dfrac{\partial B_m}{\partial R} - \left(1+\dot{\phi}\right) \dfrac{R}{F} \dfrac{\partial F}{\partial R} - \left(1+\dot{\phi}\right) \tan(\phi) \dfrac{1}{F} \dfrac{\partial F}{\partial \theta} \label{Eq:qytheta}\\
f(\theta) &=& \dfrac{\ddot{\phi}}{2} \tan(\phi)  + \dfrac{\dot{\phi}^2-1}{4} \tan ^2(\phi) + \dfrac{\dot{\phi}}{2} \left(\dot{\phi}-1 \right)
\end{eqnarray} 
\end{widetext}
Note the analogy between Eq. (\ref{Eq:qytheta}) and Eq. (5.12) in Symon's paper \cite{symon} where all the derivations are based on generalized azimuthal coordinates: the first term represents the defocusing due to the average field index of the magnets. The spiral focusing comes predominantly from the second term as is discussed in the appendix. The third term is usually small except near the edge of the magnet hence it accounts for the edge focusing effect (often called Thomas focusing). Finally, the additional term in Eq. (\ref{Eq:qxtheta}) comes from the horizontal restoring force. \\
Note in addition that the flutter function $F(R,\theta)$ can vanish in the straight sections thus yielding a singularity in the above equations. However, one can observe that the sign of the curvature function is related to the sign of the flutter function. Therefore, one shall seek a relationship between these two quantities.   

\subsubsection{Flutter function}
\begin{figure}
\centering 
\includegraphics*[width=8cm]{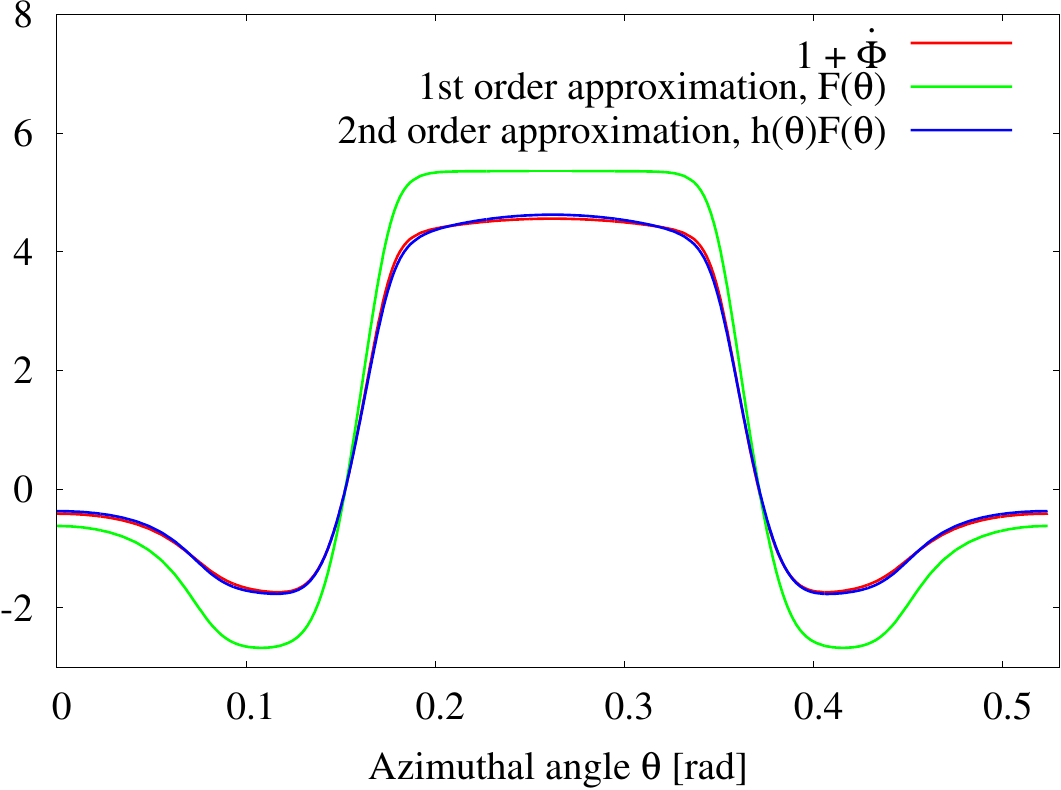}
\caption{Comparison of the tracking results (red) of the closed orbit scalloping at the KURNS FFA injection energy with its first and second approximate expressions (green and blue respectively).}
\label{fig:flutter_vs_orbit}
\end{figure}

First, let's compute the average magnetic field over a closed orbit $\mathcal{R}$ corresponding to a particle momentum $p$:
\begin{eqnarray}
\mean{B}_{co} = \dfrac{1}{2\pi \mathcal{R}} \int_0^{2\pi \mathcal{R}} \dfrac{p}{q} \dfrac{ds}{\rho(s)} = \dfrac{p}{q \mathcal{R}}
\end{eqnarray}
where $\mean{}_{co}$ refers to the average taken over the curvilinear abscissa. It results that
\begin{eqnarray}
B\rho = \dfrac{p}{q} = \mean{B}_{co} \mathcal{R}
\label{Eq:brho}
\end{eqnarray}
For the case of a radial sector FFA, the flutter function is independent of the radius, and one can write the expression of the magnetic field seen by the particle as a function of the azimuthal angle:
\begin{eqnarray}
B(R(\theta),\theta)=B_m(R(\theta)) F(\theta)
\end{eqnarray}
Now, equating the expression of the magnetic rigidity in both representations and making use of \mbox{Eqs. (\ref{Eq:rho}), (\ref{Eq:transf}), (\ref{Eq:brho}),} one finally obtains:
\begin{eqnarray}
F(\theta) &=& \dfrac{\mean{B}_{co}}{B_m(R(\theta))} \dfrac{\mathcal{R}}{R(\theta)} \left(1+\dot{\phi}\right) \cos(\phi)  \label{Eq:F_phi_ex} \\
&=& \dfrac{\mean{B}_{co}}{B_m(R(\theta))} \dfrac{d\theta}{d\vartheta} \left(1+\dot{\phi}\right) \nonumber
\end{eqnarray}
Thus, when assuming that the closed orbit does not depart much from a fixed radius orbit i.e. $\phi \ll 1$, which is generally the case for fixed field accelerators, one can write as a first order approximation:
\begin{eqnarray}
\dot{\phi} \approx F(\theta)-1 \label{Eq:F_phi}
\end{eqnarray}
Nevertheless, the above approximation becomes less valid when the orbit scalloping becomes important and when the radial increase of the field becomes large. For instance, for the KURNS FFA, the above formula is tested at injection energy where the orbit scalloping \mbox{$\Delta R/\mean{R}$} is about 2\%, and one can observe in fig. \ref{fig:flutter_vs_orbit} a non negligible difference between the two expressions. Thus, a second order approximation is established (see \mbox{appendix \ref{appendix:b})} which is more accurate to explain the azimuthal variations of the scalloping angle as illustrated in fig. \ref{fig:flutter_vs_orbit}.

\section{Number of betatron oscillations from the second order differential equation}
As was established by Teng \cite{teng}, if one defines $K_{x,y}(\theta)$ as the forcing term of the Hill's equation \mbox{($K_{x,y}(\theta)=q_{x,y}(\theta)+f(\theta)$),} then the number of betatron oscillations can be approximated by:
\begin{eqnarray}
\nu_i^2 \approx \mean{K_i} &+& \mean{\widetilde{K_i}^2} \nonumber \\
&+& 3 \mean{K_i}\mean{\widetilde{\widetilde{K_i}}^2} + \mean{K_i \widetilde{\widetilde{K_i}}^2}
\label{Eq:tune_BKM}
\end{eqnarray}
where the subscript $i$ stands for $x$ or $y$, the symbol $\mean{}$ represents the average taken over the azimuthal coordinate and the tilde \hspace{1mm}$\widetilde{•}$ \hspace{0.3mm} is the integrating operator defined by: 
\begin{eqnarray}
\widetilde{g}(\theta) = \int_{}^{} \left[ g(\theta)-\mean{g} \right] d\theta  \nonumber \\
\widetilde{\widetilde{g}}(\theta) = \int_{}^{} \left[ \widetilde{g}(\theta)-\mean{\widetilde{g}} \right] d\theta  \nonumber
\end{eqnarray}
For each closed orbit, there is a different set of linearized equations for the betatron oscillations, i.e. different pairs $(K_x(\theta),K_y(\theta))$ so that the betatron wave numbers are susceptible to change with the energy.    
\\
For a radial sector FFA, when the closed orbit scalloping and the field gradients are moderate, the Alternating Gradient (AG) effect can be neglected. in other words the tunes can be well approximated by the first order term in Eq. (\ref{Eq:tune_BKM}) which is simply the average of the forcing term of the second order linear differential equation (\ref{Eq:qx}). Nevertheless, when the $k$-value increases substantially, such an approximation no longer holds and one needs to account for the higher order terms which are due to the AG forces that produce a substantial scalloping of the orbits alongside a stronger focusing. This is accounted for by the tilde functions.\\
In what follows, the different terms of the Hill's equation expressed in cylindrical coordinates will be discussed and the emphasis made on their contribution to the number of betatron oscillations. Several formula will be established in the limit where the orbit scalloping is small i.e. relying on the first order approximation (Eq. (\ref{Eq:F_phi})). To conclude, one will make a comparison between the tracking results and the analytical estimates for various values of the FD ratio and the average field index of the magnets.

\subsection{Focusing due to the average field index}
If the averaged field (over the azimuth) increases with the radius, this will yield an overall focusing (resp. defocusing) force in the horizontal (resp. vertical) plane. Such a contribution is defined by:
\begin{eqnarray}
q_x^{ind}(\theta) = \left(1+\dot{\phi}\right) \dfrac{R}{B_m} \dfrac{\partial B_m}{\partial R} = \left(1+\dot{\phi}\right) k
\end{eqnarray}
Next, two concepts will be discussed based on the expression of the average magnetic field.

\subsubsection{Cyclotron}
In order to keep the isochronism in a cyclotron, the average magnetic field strength changes according to the law:
\begin{eqnarray}
B_m(R) = B_0 \gamma(R) = \dfrac{B_0}{\sqrt{1-\left(\dfrac{R}{R_{\infty}} \right)^2}} 
\end{eqnarray}
where $R_{\infty}$ is a constant. It results that:
\begin{eqnarray}
\mean{q_x^{ind}} = k = \gamma^2 - 1 
\end{eqnarray}
which shows that, in general, the average field index contributes weakly to the horizontal focusing in cyclotrons. Such a weak contribution is advantageous in the vertical plane since it can be overcome by means of Thomas and/or spiral focusing. 

\subsubsection{Scaling FFA}
Assuming that $k$ is constant everywhere in the ring, i.e. for a perfect scaling FFA, it results that:
\begin{eqnarray}
\mean{q_x^{ind}} &=& k = \text{const} \label{Eq:qxk}
\end{eqnarray}
Since the $k$-value can be large, $k=7.6$ for the KURNS scaling FFA, it is obvious that the higher order terms in \mbox{Eq. (\ref{Eq:tune_BKM})} which account for the AG forces can no longer be neglected. \\
For instance, for a machine like the KURNS 150 MeV FFA where $k=7.6$, one can estimate the contribution of the AG due to the field index to be:
\begin{eqnarray}
\mean{\widetilde{q_x^{ind}}^2} &=& k^2 \mean{\phi^2}  \nonumber \\
&\approx & 7.6^2 \mean{(0.33*\sin(12 \theta))^2} = 3.14 \label{Eq:qxtilde_kurri}
\end{eqnarray} 
which is about 2/5 of the focusing due to the average field index. This is non negligible and can place the tunes in the second stability region of Hill's equation, hence the concept of scaling FFA with small orbit excursion \cite{small_orbit_excursion}. 
However, this will be discussed more in detail later on in this paper where the AG focusing is evaluated.

\subsection{Thomas focusing}
Under the assumption of small orbit scalloping, \mbox{Eq. (\ref{Eq:F_phi})} holds and  the contribution of the Thomas focusing to the vertical tune can be simplified to:
\begin{eqnarray}
q_y^{th}(\theta)= - \left(1+\dot{\phi}\right) \tan(\phi) \dfrac{1}{F} \dfrac{\partial F}{\partial \theta} \approx - \phi \ddot{\phi}
\end{eqnarray}
By means of an integration by parts and making use of Eq. (\ref{Eq:F_phi}), one finally obtains:
\begin{eqnarray}
\mean{q_y^{th}} \approx \mean{\dot{\phi}^2} \approx \mean{\left(F(\theta)-1\right)^2} = \mathcal{F}^2 \label{Eq:meanqyth}
\end{eqnarray}
where $\mathcal{F}^2$ is generally referred to as the magnetic flutter and can also be re-written in the following form:
\begin{eqnarray}
\mathcal{F}^2 = \dfrac{\mean{F^2}-\mean{F}^2}{\mean{F}^2} = \dfrac{\mean{B^2}-\mean{B}^2}{\mean{B}^2}
\end{eqnarray}
The latter represents the fractional mean square azimuthal deviation of the field at a fixed radius \cite{craddock}. This term is large when the field is changing rapidly along the closed orbit.\\
In general, Thomas focusing is dominated by the edge focusing effect and acts in a way that the resulting effect is a net restoring force in the vertical plane. It was Thomas in his seminal paper of 1938 \cite{thomas} who showed for the first time that using an azimuthally varying field producing a scalloped particle orbit allows to overcome the defocusing effect that would set the maximum energy of cyclotrons to 20 MeV as claimed by Bethe and \mbox{Rose \cite{bethe}.} \\
Thomas's insight was to realize that the interaction between the radial component of the momentum and the azimuthal component of the magnetic field yields a vertical restoring force:
\begin{eqnarray}
F_y = -q v_R B_{\theta} = -q \dfrac{dR}{d\theta}\dfrac{d\theta}{dt} B_{\theta} \label{Eq:Fthomas}
\end{eqnarray}
where $B_{\theta}$ is the azimuthal component of the magnetic field which writes for small $z$ \cite{thomas}:
\begin{eqnarray}
B_{\theta} = \dfrac{z}{R}\dfrac{\partial B}{\partial \theta} + O(z^3) \label{Eq:btheta}
\end{eqnarray}
Injecting the latter into the expression of $F_y$ and making use of Eq. (\ref{Eq:tan}) yields:
\begin{eqnarray}
F_y = q \dfrac{d\theta}{dt} \tan(\phi) \dfrac{\partial B}{\partial \theta} z
\end{eqnarray}
Such a force is usually directed towards the mid-plane and therefore it is restoring in the vertical plane. Its average contribution to the vertical tune is approximately given by Eq. (\ref{Eq:meanqyth}) which is mainly valid in the limit of small scalloping of the orbit \mbox{i.e. $\phi \ll 1$.}
\begin{figure}
\centering 
\includegraphics*[width=8cm]{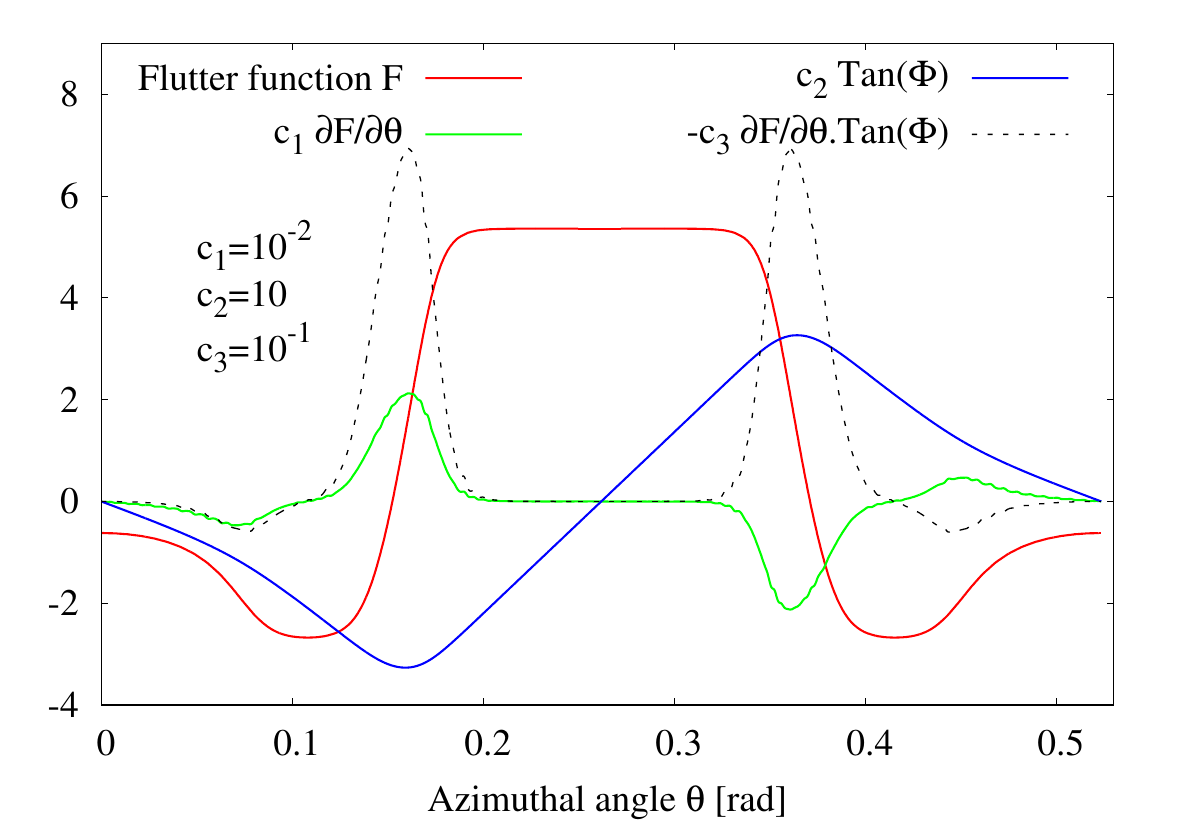}
\caption{Plot of the different terms contributing to the Thomas focusing at the injection energy of the KURNS scaling FFA: the orbit scalloping is shown via the angle $\phi$ (in blue) while the contribution of the field index is shown in green (partial derivative of the flutter \mbox{function $F$.}) Their product is shown in dashed lines which accounts for the vertical focusing. }
\label{fig:thomas_foc}
\end{figure}
This is further summarized in fig. \ref{fig:thomas_foc} displaying the tracking results from the KURNS 150 MeV FFA. 

\subsection{Horizontal restoring force}

Larger scalloping of the orbit is due to larger oscillations of the flutter function $F$ which can substantially increase the horizontal restoring force:
\begin{eqnarray}
\mean{q_x^{res}} = \dfrac{1}{2\pi} \int_0^{2\pi} \left(1+\dot{\phi}\right)^2 d\theta &=& 1 + \mean{{\dot{\phi}}^2} \nonumber \\
 &\approx & 1 + \mathcal{F}^2
\end{eqnarray}
It results that, for the horizontal tune, the horizontal restoring force and the Thomas defocusing term compensate each other in such a way that: 
\begin{eqnarray}
\mean{q_x^{res}} - \mean{q_y^{th}} \approx 1 \label{Eq:qxres_qyth}
\end{eqnarray}
In addition, stronger vertical focusing can be achieved if the magnet boundaries are deformed from radial poles to spiral shaped poles. This is further discussed in the appendix \ref{appendix:a}. However, for the remaining part of the present paper, only radial sector machines are considered so that such a contribution is disregarded.

\subsection{Alternating Gradient focusing}
From the previous analysis, and assuming small orbit scalloping to simplify the calculations, one can write the expression of $K_x$ for a radial sector scaling FFA:
\begin{eqnarray}
K_x(\theta) &=& q_x^{res}(\theta) + q_x^{ind}(\theta) - q_y^{th}(\theta) + f(\theta) \nonumber \\
& \approx & \left(1 + \dot{\phi} \right)^2 +  \left(1 + \dot{\phi} \right)k + \dfrac{3}{2} \phi \ddot{\phi} + \dfrac{\dot{\phi}^2}{2} -\dfrac{\dot{\phi}}{2} \nonumber
\end{eqnarray}
where one neglected a small term involving $f(\theta)$ so that $\mean{K_x} \approx k+1$, and
\begin{eqnarray}
\widetilde{K_x}(\theta) &=& \int_0^{\theta} \left( K_x(u) - \mean{K_x} \right) du  \nonumber \\
& \approx & \left[k+\dfrac{3}{2} (1+\dot{\phi}) \right] \phi  \text{  ;  } \phi(0)=0 \label{Eq:Kx_exact}
\end{eqnarray}
Similarly, one can compute the contribution of the AG to the vertical focusing:
\begin{eqnarray}
K_y(\theta) &=& - q_x^{ind}(\theta) + q_y^{th}(\theta) + f(\theta) \nonumber \\
& \approx & - \left(1 + \dot{\phi} \right)k - \dfrac{\phi \ddot{\phi}}{2} + \dfrac{\dot{\phi}^2}{2} -\dfrac{\dot{\phi}}{2} \nonumber
\end{eqnarray}
which yields after integration $\mean{K_y} \approx -k+\mean{\dot{\phi}^2}$ and
\begin{eqnarray}
\widetilde{K_y}(\theta) &=& \int_0^{\theta} \left( K_y(u) - \mean{K_y} \right) du  \label{Eq:Ky_exact} \\
& \approx & -\left[k+\dfrac{1}{2} (1+\dot{\phi}) \right] \phi + \int_0^\theta \left[ \dot{\phi}^2 - \mean{\dot{\phi}^2} \right] du \nonumber
\end{eqnarray}

To facilitate further discussion, one makes the following considerations: let's assume that $F(\theta)$ is an even function in $\theta$ as illustrated in fig. \ref{fig:flutter_vs_orbit}. It results that its integral and therefore the scalloping angle $\phi$ as given by Eq. (\ref{Eq:F_phi}) is an odd function in $\theta$. Thus, $\phi$ writes in the form:
\begin{eqnarray}
\phi(\theta) = \sum_{j=1}^{\infty} \phi_j \sin \left(j N \theta \right) 
\end{eqnarray}
where $\phi_{j}$ are the coefficients of the Fourier series and $N$ is the total number of sectors in the ring.
To simplify the analysis, one shall keep only the first two terms in this series:
\begin{eqnarray}
\phi(\theta) = \phi_{1} \sin(N\theta) + \phi_{2} \sin(2N\theta) \label{Eq:phi_series}
\end{eqnarray}
Now, making use of Eqs. (\ref{Eq:Kx_exact}) and (\ref{Eq:Ky_exact}) one obtains:
\begin{eqnarray}
\mean{\widetilde{K_x}^2} = \left(k+\dfrac{3}{2}\right)^2 && \dfrac{{\phi_{1}}^2 + {\phi_{2}}^2}{2} + \dfrac{9N^2}{32} \left({\phi_1}^4 + 4 {\phi_2}^4 \right) \nonumber \\
& & + \dfrac{45}{16} N^2 {\phi_1}^2 {\phi_2}^2 \label{Eq:kxtilde2}
\end{eqnarray}
and
\begin{eqnarray}
\mean{\widetilde{K_y}^2} = \left(k+\dfrac{1}{2}\right)^2 && \dfrac{{\phi_{1}}^2 + {\phi_{2}}^2}{2} + \dfrac{365}{144} N^2 {\phi_1}^2 {\phi_2}^2 \nonumber \\
& & - \dfrac{9}{4} \left(k+\dfrac{1}{2} \right) N {\phi_1}^2 {\phi_2} \label{Eq:kytilde2}
\end{eqnarray}
If $\phi_2=0$, then one can write an approximate expression of the tunes as follows:
\begin{eqnarray}
\nu_x &\approx & \left[k+1 + \left(k+\dfrac{3}{2}\right)^2 \mean{\phi^2} + \dfrac{9}{8}N^2 \mean{\phi^2}^2 \right]^{1/2} \\
\nu_y &\approx & \left[-k+\mean{\dot{\phi}^2} + \left(k+\dfrac{1}{2}\right)^2 \mean{\phi^2} \right]^{1/2}
\end{eqnarray}
where one assumed that most of the focusing comes predominantly from the first two terms in the tune expression (\ref{Eq:tune_BKM}) and where the scalloping angle $\phi$ is obtained from its first \mbox{($\phi \approx \widetilde{F(\theta)}$)} or improved second order approximation \mbox{($\phi \approx \widetilde{h(\theta)F(\theta)}$)}. \\
The above expressions of the tunes show that the AG effect is generally more important in the horizontal plane than in the vertical one. The latter is particularly sensitive to the shape of the closed orbit. In addition, the contribution of the AG to the transverse focusing increases with the number of sectors.  \\
Now, as a verification example, let's fit the tracking data of the scalloping angle $\phi$ at the injection energy of the KURNS 150 MeV FFA with the form given by Eq. (\ref{Eq:phi_series}). This yields $\phi_1=-0.2557$ rad and $\phi_2=0.0673$ rad. Injecting the latter into Eqs. (\ref{Eq:kxtilde2}) and (\ref{Eq:kytilde2}) yields:
\begin{eqnarray}
\nu_x \approx \left[ \mean{K_x} + \mean{\widetilde{K_x}^2} \right]^{1/2} = \left[ 7.84 + 3.43 \right]^{1/2} = 3.36 \nonumber \\
\nu_y \approx \left[ \mean{K_y} + \mean{\widetilde{K_y}^2} \right]^{1/2} = \left[ 0.92 + 1.44 \right]^{1/2} = 1.53 \nonumber
\end{eqnarray}
which is consistent with the tracking simulation results using the TSOCA 3D field map as shown in fig. \ref{fig:tunes_meas}. In addition, it is clear that the AG effect is negligible for radial sector fixed field machines where the $k$-value is not important such as cyclotron accelerators or for machines where the orbit scalloping is small i.e. $F(\theta) \approx 1$. When the $k$-value is non negligible as well as the orbit scalloping, which is the focus of our analysis, then the net focusing is due to a mixing between the average forces applied on the beam and the alternating sign of these forces.

\subsection{Benchmarking the analytical formula with tracking simulations} \label{subsection:benchm}
In summary, the number of betatron oscillations for a radial sector scaling FFA can be determined from the expressions of $K_x$ and $K_y$:
\begin{eqnarray}
K_x(\theta) &=&  q_x^{res}(\theta) + q_x^{ind}(\theta) - q_y^{th}(\theta) + f(\theta) \nonumber \\
K_y(\theta) &=&  -q_x^{ind}(\theta) + q_y^{th}(\theta) + f(\theta) \nonumber
\end{eqnarray}
where $f(\theta)$ is given by Eq. (\ref{Eq:qy}) such that its first order contribution is a small defocusing effect in both planes:
\begin{eqnarray}
\mean{f} = -\dfrac{\mean{p^2}}{4} = - \dfrac{1}{4} \mean{(1-\dot{\phi})^2 \tan^2(\phi)}
\end{eqnarray}
and the tunes are evaluated by means of Eq. (\ref{Eq:tune_BKM}). \\
Table \ref{table1} shows a comparison of the tracking results of radial sector scaling FFA with the analytical expressions for different values of the average field index: as can be seen, when increasing the $k$-value, the vertical tune decreases. However, this is explained by two main contributions: an increasing average field index as well as a decreasing edge focusing. The latter is particularly sensitive to the shape of the equilibrium orbit which depends on the average field index as well as the flutter function. In addition, one can see that the often quoted formula $\nu_x^2 \approx k+1$ and $\nu_y^2 \approx -k+\mathcal{F}^2$ do predict the qualitative behavior. However, they fail to predict the quantitative (or monotonic behavior as will be shown later on in this paper). In general, for small $k$-values, one can observe that the vertical tunes are well approximated by the average values of the transverse forces applied on the beam. This is the cyclotron regime for which the average field index does not depart much from $\gamma^2-1$. Nevertheless, the horizontal tunes are not well predicted by such an approximation. The reason is due to the FD ratio which is defined as the absolute value of the ratio between the minimum and maximum of the flutter function i.e. $\text{FD}=|\text{min}(F)/\text{max}(F)|$. The latter is non negligible in this example, FD$=0.69$,  so that the AG forces dominate in the horizontal plane. Hence the need to account for the additional focusing terms given by Eq. (\ref{Eq:tune_BKM}). This is further illustrated in table \ref{table2} where one can see that the discrepancy between the tracking and the 1st order formula increases with increasing FD ratio. Note that the edge focusing term was calculated from its exact expression so that the effect of the orbit scalloping which depends on the average field index $k$ as established in appendix \ref{appendix:b} is precisely evaluated. Nevertheless, it appears that the smooth approximation fails near the stability boundary hence the need to extend the approximation to higher order terms \cite{teng}. However, we will not pursue this here.

\begin{table*}[]

  \centering
  \resizebox{1.0\textwidth}{!}{\begin{minipage}{\textwidth}
        \caption{Comparison of the zgoubi tracking results with the approximate formula for various values of the average field \mbox{index $k$.}}

\begin{tabular*}{1.0\textwidth}%
   {@{\extracolsep{\fill}}||l|ll|ll|ll|l||}
\hline \hline
\multicolumn{1}{|c|}{\multirow{2}{*}{\begin{tabular}[c]{@{}c@{}}Field \\ index k\end{tabular}}} & \multicolumn{2}{c|}{Tracking}                                                           & \multicolumn{2}{c|}{1st order}                                                          & \multicolumn{2}{c|}{3rd order}                                                          & \multicolumn{1}{c|}{\multirow{2}{*}{\begin{tabular}[c]{@{}c@{}}Edge focusing\\ $\mean{q_y^{th}}$ \end{tabular}}} \\ \cline{2-7}
\multicolumn{1}{|c|}{}                                                                        & \multicolumn{1}{c|}{$\nu_x^2$} & \multicolumn{1}{c|}{$\nu_y^2$} & \multicolumn{1}{c|}{$\mean{K_x}$} & \multicolumn{1}{c|}{$\mean{K_y}$} & $\mean{K_x} + \mean{\widetilde{K_{x}}^2} + ...$ & $\mean{K_y} + \mean{\widetilde{K_{y}}^2} + ...$ & \multicolumn{1}{c|}{}                               \\ \hline \hline
0                                                                                             & 1.40                                       & \textbf{22.89}                             & (-1.27)                                    & 21.72                                      & 1.17                                       & \textbf{23.02}                             & 22.02                                               \\ 
1                                                                                             & 2.97                                       & \textbf{17.68}                             & 0.42                                       & 17.41                                      & 2.89                                       & \textbf{17.73}                             & 18.68                                               \\ 
2                                                                                             & 4.67                                       & \textbf{14.25}                             & 1.86                                       & 14.13                                      & 4.67                                       & \textbf{14.22}                             & 16.31                                               \\ 
3                                                                                             & 6.54                                       & \textbf{11.67}                             & 3.17                                       & 11.43                                      & 6.61                                       & \textbf{11.66}                             & 14.58                                               \\ 
4                                                                                             & 8.59                                       & \textbf{9.62}                              & 4.38                                       & 9.09                                       & 8.72                                       & \textbf{9.62}                              & 13.20                                               \\ 
5                                                                                             & 10.86                                      & \textbf{7.89}                              & 5.56                                       & 7.02                                       & 11.02                                      & \textbf{7.87}                              & 12.10                                               \\ 
6                                                                                             & 13.43                                      & \textbf{6.40}                              & 6.70                                       & 5.12                                       & 13.51                                      & \textbf{6.30}                              & 11.20                                               \\ 
7                                                                                             & 16.45                                      & \textbf{5.07}                              & 7.81                                       & 3.36                                       & 16.23                                      & \textbf{4.81}                              & 10.44                                               \\ 
8                                                                                             & 20.19                                      & \textbf{3.90}                              & 8.91                                       & 1.73                                       & 19.13                                      & \textbf{3.35}                              & 9.80                                                \\ 
9                                                                                             & 25.55                                      & \textbf{2.80}                              & 9.91                                       & 0.17                                       & 22.25                                      & \textbf{1.86}                              & 9.23                                                \\ 
10                                                                                            & 32.01                                      & \textbf{1.86}                              & 11.06                                      & (-1.32)                                    & 25.58                                      & \textbf{0.32}                              & 8.75                                                \\ \hline \hline
\end{tabular*}
\label{table1}
\end{minipage}}

\end{table*}

\begin{table*}[]

  \centering
  \resizebox{1.0\textwidth}{!}{\begin{minipage}{\textwidth}
        \caption{Comparison of the zgoubi tracking results with the approximate formula for various FD ratio ($k=5$).}

\begin{tabular*}{1.0\textwidth}%
   {@{\extracolsep{\fill}}||l|ll|ll|ll|l||}
\hline \hline
\multicolumn{1}{|c|}{\multirow{2}{*}{\begin{tabular}[c]{@{}c@{}}FD \\ ratio\end{tabular}}} & \multicolumn{2}{c|}{Tracking}                                                           & \multicolumn{2}{c|}{1st order}                                                          & \multicolumn{2}{c|}{3rd order}                                                          & \multicolumn{1}{c|}{\multirow{2}{*}{\begin{tabular}[c]{@{}c@{}}Edge focusing\\ $\mean{q_y^{th}}$ \end{tabular}}} \\ \cline{2-7}
\multicolumn{1}{|c|}{}                                                                        & \multicolumn{1}{c|}{$\nu_x^2$} & \multicolumn{1}{c|}{$\nu_y^2$} & \multicolumn{1}{c|}{$\mean{K_x}$} & \multicolumn{1}{c|}{$\mean{K_y}$} & $\mean{K_x} + \mean{\widetilde{K_{x}}^2} + ...$ & $\mean{K_y} + \mean{\widetilde{K_{y}}^2} + ...$ & \multicolumn{1}{c|}{}                               \\ \hline \hline
0.34                                                                                             & 7.88                                       & \textbf{0.81}                             & 5.97                                    & 0.48                                      & 8.14                                       & \textbf{0.78}                             & 5.51                                               \\ 
0.47                                                                                             & 8.73                                       & \textbf{3.00}                             & 5.89                                       & 2.52                                      & 9.02                                       & \textbf{2.97}                             & 7.57                                               \\ 
0.61                                                                                             & 10.00                                       & \textbf{6.00}                             & 5.71                                       & 5.30                                      & 10.25                                       & \textbf{5.99}                             & 10.37                                               \\ 
0.75                                                                                             & 11.93                                       & \textbf{10.43}                             & 5.36                                       & 8.96                                      & 11.90                                       & \textbf{10.02}                             & 14.07                                               \\ 
0.89                                                                                             & 14.99                                       & \textbf{15.36}                              & 4.74                                       & 13.58                                       & 14.02                                       & \textbf{15.24}                              & 18.78                                               \\ 
1.10                                                                                             & 25.69                                      & \textbf{27.60}                              & 3.06                                       & 22.42                                       & 18.15                                      & \textbf{25.54}                              & 27.86                                                \\ \hline \hline
\end{tabular*}
\label{table2}
\end{minipage}}

\end{table*}

\section{Beam stability analysis}
Technically, it is impossible to make a field which corresponds exactly to the designed one. Therefore, it is important to understand the effect of small imperfections of the field on the beam dynamics. \\
The general equations of motion including non-linear terms and imperfections are defined by:
\begin{eqnarray}
\begin{cases}
& \dfrac{d^2x}{ds^2}=P(x,y,s)   \\  \\
& \dfrac{d^2y}{ds^2}=Q(x,y,s)
\label{non_lin_eq}
\end{cases}
\end{eqnarray} 
where $x$ and $y$ represent the deviation around the closed orbit, $s$ the curvilinear coordinate and $P$ and $Q$ are real analytic functions of $x$, $y$ \mbox{and $s$.} \\
In the following, we investigate the stability of the particle trajectories that can arise with different field errors. We use two different approaches to investigate the beam stability due to field errors: the first approach is based on the previously established analytical solution of the betatron wave numbers by means of a smooth approximation of the linearized equations of motion. The second model is the zgoubi \cite{zgoubi} tracking model of the magnet, which is the most accurate one. The zgoubi model solves the non-linear equation of motion using field maps or user-implemented analytical models. To conclude, we establish a comparison between the different results and comment on the outcome of this study.

\subsection{Field imperfections in scaling FFAs}
\begin{figure}
\centering 
\includegraphics*[width=8cm]{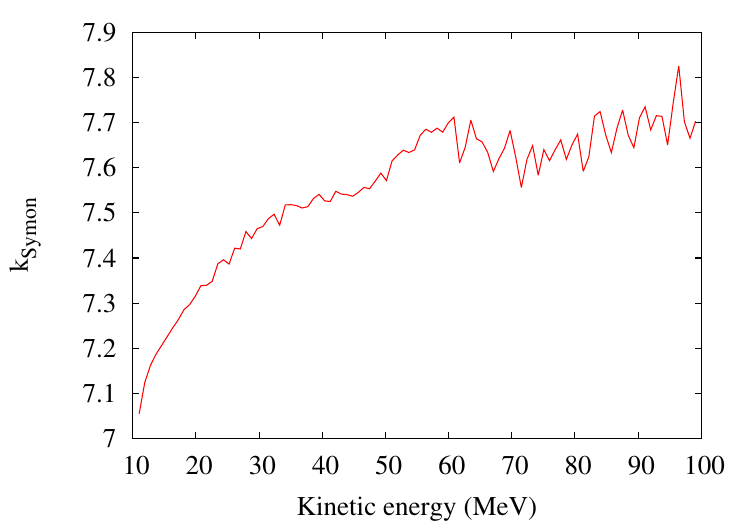}
\caption{Plot of the average field index $k_{symon}$ as a function of the kinetic energy for the KURNS 150 MeV scaling FFA. Note the oscillatory behavior which is due to the granularity of the field map.}
\label{fig:ksymon}
\end{figure}

For an ideal radial sector scaling FFA, the magnetic field writes in cylindrical coordinates in the following way:
\begin{eqnarray}
B(R,\theta) = B_0 \left(\dfrac{R}{R_0} \right)^k F(\theta) \label{Eq:scaling_FFA}
\end{eqnarray}
Thus, a natural way to verify the validity of a calculated field map is to introduce a generalized definition of the average field index which accounts for the local imperfections of the field (radially and azimuthally): 
\begin{eqnarray}
k(R,\theta) = \dfrac{R}{B} \dfrac{\partial B}{\partial R}
\end{eqnarray}
where one shall exclude the field-free region from the definition of $k(R,\theta)$. \\
In order to avoid any confusion, it is important to remind the definition of the average field index introduced by Symon which is given by:
\begin{eqnarray}
k_{symon} = \dfrac{\mathcal{R}}{\mean{B}_{co}} \dfrac{d\mean{B}_{co}}{d\mathcal{R}}
\end{eqnarray}
If Eq. (\ref{Eq:scaling_FFA}) holds everywhere in the ring, then \mbox{$k(R,\theta)=k_{symon}$} is constant and both definitions are equivalent. Nevertheless, due to imperfections one expects some variations of the above defined quantities as illustrated in fig. \ref{fig:ksymon}. 

\subsubsection{Field map derivative}
\begin{figure}
\centering 
\includegraphics*[width=8cm]{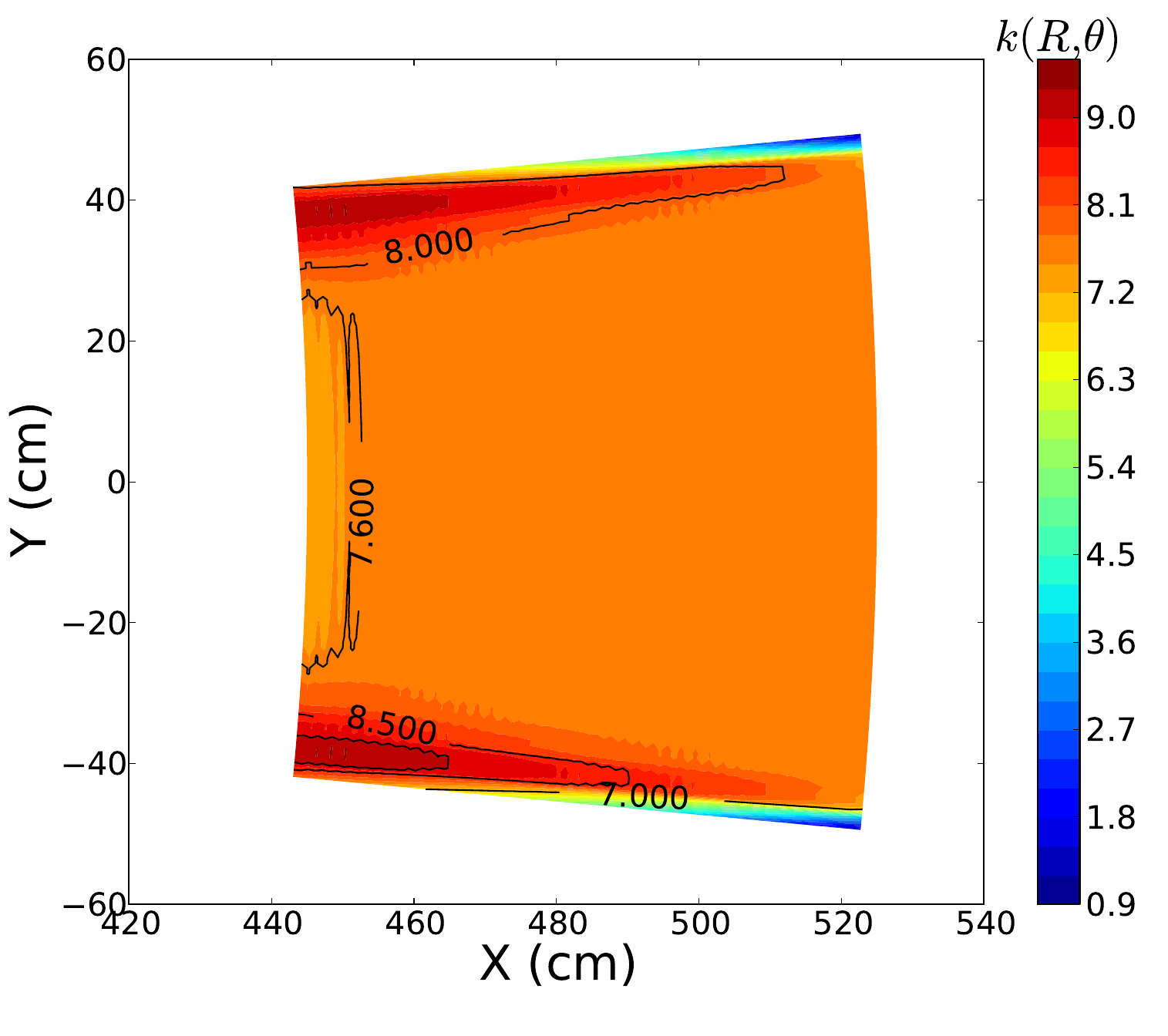}
\caption{Plot of the average field index map of the F-magnet. The central line ($Y=0$) is a line of symmetry of the DFD triplet. }
\label{fig:kF_map}
\end{figure}

\begin{figure}
\centering 
\includegraphics*[width=8cm]{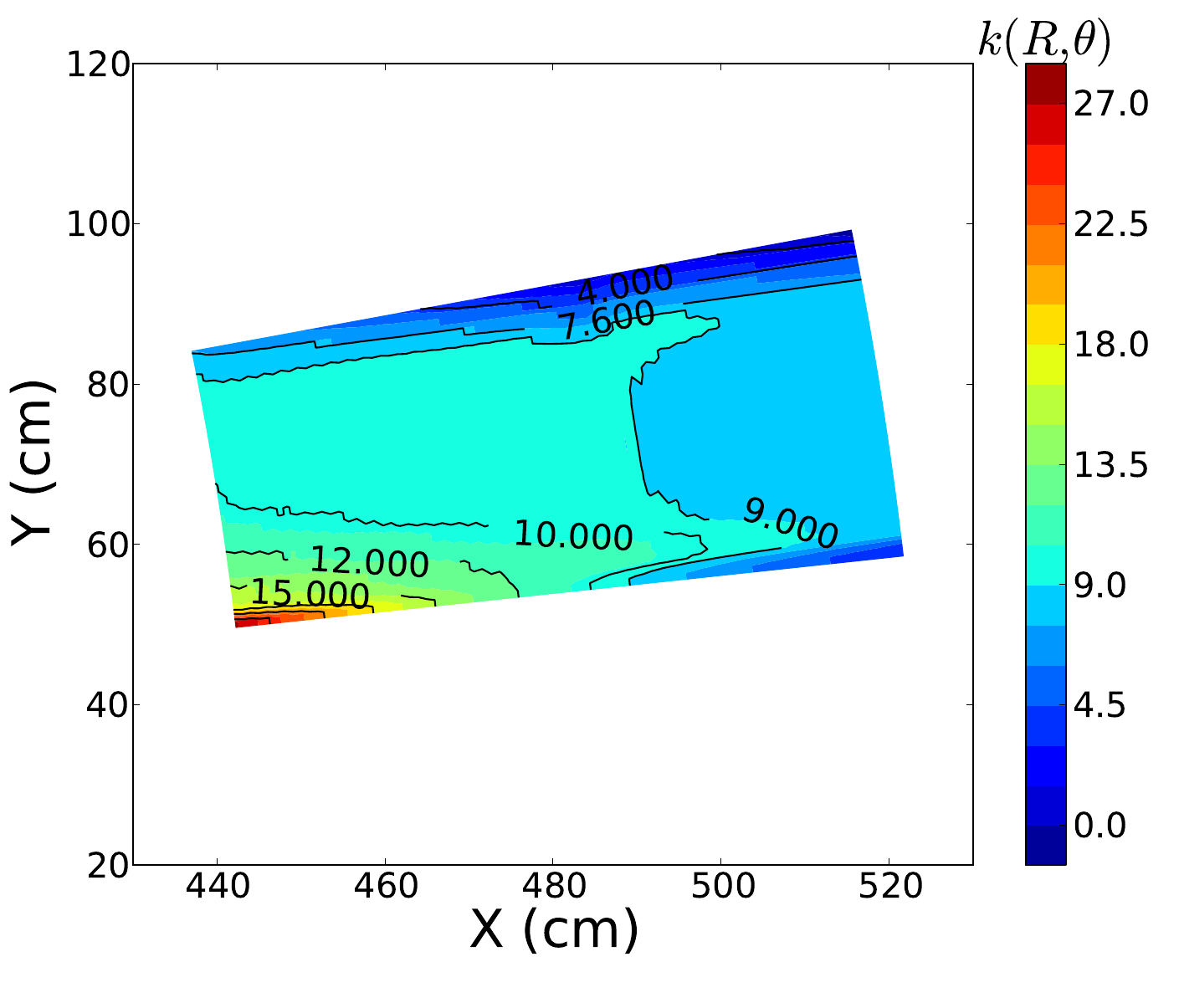}
\caption{Plot of the average field index map of the D-magnet. The lower part of the magnet is in the neighborhood of the F-magnet while the upper part is surrounded by the drift space separating the sectors. }
\label{fig:kD_map}
\end{figure}

\begin{figure}
\centering 
\includegraphics*[width=8cm]{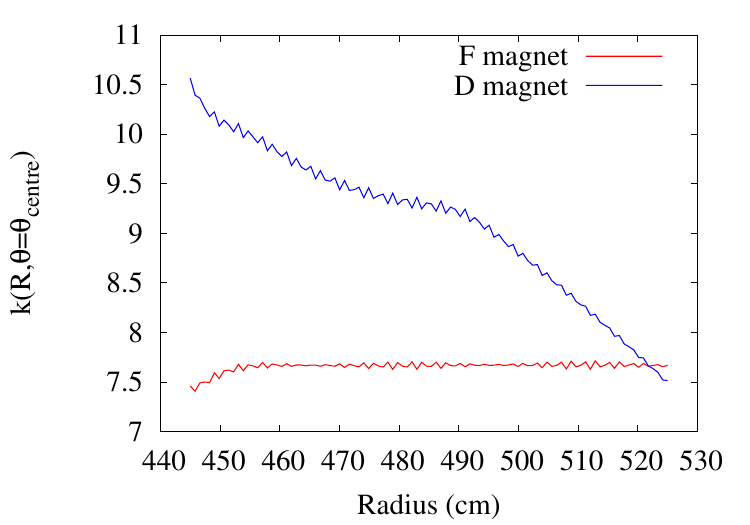}
\caption{$k(R,\theta)$ in the centre of the F and D magnet. }
\label{fig:kcentral}
\end{figure}

Using TOSCA 2D median plane field maps of the KURNS FFA, one calculated the field map derivative $k(R,\theta)$ of the main magnets, i.e. the focusing F-magnet and the defocusing D one. The result is shown in \mbox{figs. \ref{fig:kF_map} and \ref{fig:kD_map}} where one can observe a non negligible variation of this quantity: for the F-magnet, the variations of $k$ are small and the latter is close to its design value $k=7.6$. Nevertheless, for the D-magnet, the variations of $k$ are important. The main source of discrepancy is observed in the interaction region between the two magnets and seems to affect mainly the defocusing one in the neighborhood of the injection radius. Each magnet is comprised of a DFD triplet configuration which is symmetrical around the center of the F magnet. Besides, a "return-yoke-free" design has been developed to ease the problem of variable energy extraction. This means that the flux generated by the F-poles return through the D-poles \cite{adachi,ishi18}. Nevertheless, this causes a non negligible leakage field in the straight section and affects mainly the D magnet. In order to have a better understanding of the impact of such a discrepancy, one can simplify the model by assigning an average field index to each magnet. The latter is calculated in the central region of each and evolves with the radius as shown in fig. \ref{fig:kcentral}. 

\subsection{Generalized model of imperfect scaling radial sector FFA}

\begin{figure}[htb]
\centering 
\includegraphics*[width=8cm]{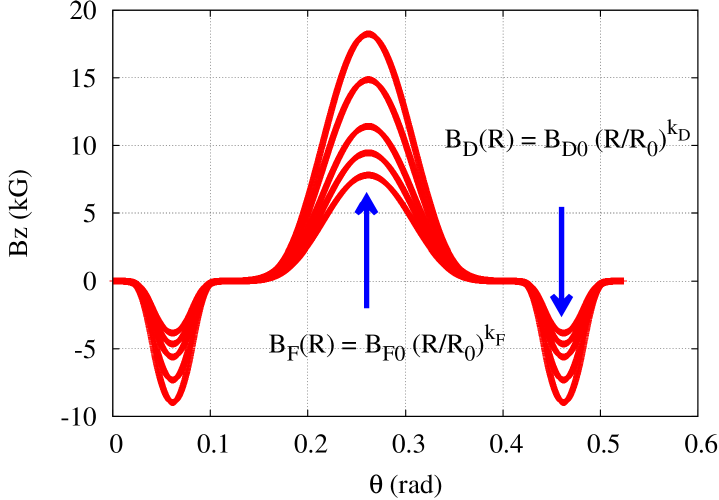}
\caption{Magnetic field along several closed orbits in a DFD triplet.}
\label{field_co}
\end{figure}

In order to carry out parametric studies of the field defects in scaling FFAs, one decided to assign an average field index to each magnet that is not necessarily equal to the ideal one. Thus, one can write: 
\begin{eqnarray}
k_i= \dfrac{R}{B_i}\dfrac{\partial B_i}{\partial R} \hspace{5mm};\hspace{5mm} i=F, D, drift \label{generalized_field_index}
\end{eqnarray}
where $B_i$ is the vertical component of the magnetic field in the median plane of the FFA magnet. Since the drift space between the magnets is likely to contain the fringe fields, it is important to assign a field index to it to determine its effect on the beam dynamics. However, in the ideal case, $k_{drift}=0$, which we will assume in the future for simplicity, yet without loss of generality. Now, assuming that the $k$-values have no radial dependence (a complete derivation of the expression of the field when $k$ is R-dependent can be found in the Appendix \ref{appendix:c}), \mbox{Eq. (\ref{generalized_field_index})} can be integrated and the magnetic field expressed in cylindrical coordinates:
\begin{eqnarray}
B(R,\theta) = &B_{F0}&  \left( \dfrac{R}{R_0} \right)^{k_F}  F_F(\theta) \nonumber \\
&+& a B_{D0} \left( \dfrac{R}{R_0} \right)^{k_D} F_D(\theta) \label{zg_model}
\end{eqnarray}  
where $F_F$ and $F_D$ are the fringe field factors (or flutter functions) that describe the azimuthal variation of the field in the $F$ and $D$ magnets respectively, and $a$ is a scale factor that allows to vary the FD ratio of the magnet ($a \ge 0$).  It is important to note that the field is a separable function in radial and azimuthal coordinates since the fringe fields merge to zero in our model between the magnets as can be seen in \mbox{fig. \ref{field_co}}, thus $F_F(\theta)F_D(\theta)=0$. Also, note that if $k_F=k_D$, the field writes in the standard form of a scaling FFA. The lattice considered for this study is a radial sector KURNS-like DFD triplet \cite{HB_suzie}.
It results that:
\begin{eqnarray}
\dfrac{1}{B} \dfrac{\partial B}{\partial \theta} &=& \dfrac{1}{F_i} \dfrac{\partial F_i}{\partial \theta} = \dfrac{1}{F} \dfrac{\partial F}{\partial \theta}  \\
\dfrac{1}{B} \dfrac{\partial B}{\partial R} &=& \dfrac{k_i(R)}{R} 
\end{eqnarray}
where $F(\theta)=F_F(\theta)+F_D(\theta)$. \\
Assuming that the average field index $k_i$ within each element evolves slowly as a function of the radius in order to neglect its radial dependence for a specific closed orbit, one finally obtains:
\begin{eqnarray}
\mean{q_x^{ind}}(p) &=& \dfrac{1}{2\pi} \int_0^{2\pi} \left(1+\dot{\phi}\right) \dfrac{R}{B} \dfrac{\partial B}{\partial R} d\theta \nonumber \\ 
&\approx & \dfrac{1}{2\pi} \sum_{i=1}^N  k_i(R) \int_{\theta_i}  (1+\dot{\phi}) d\theta \nonumber \\
&=& \sum_{i=1}^N \alpha_i(p) k_i(p) \label{Eq:qxind}
\end{eqnarray}
where $N$ is the sum of all the elements in the ring and $\alpha_i$ represents the signed fractional curvature for an element $i$ with a specific momentum $p$:
\begin{eqnarray}
\alpha_i(p) = \dfrac{1}{2\pi} \int_{\theta_i}  (1+\dot{\phi}) d\theta 
\end{eqnarray} 
therefore satisfying
\begin{eqnarray}
\sum_{i=1}^N \alpha_i(p) = 1
\end{eqnarray}
Assuming that the ring contains two types of magnets F and D, then Eq. (\ref{Eq:qxind}) becomes:
\begin{eqnarray}
\mean{q_x^{ind}} = k_F + \alpha_D(p) \left( k_D - k_F \right)
\end{eqnarray}
Given that $\alpha_D < 0$, the established equation shows that the contribution of the average field index to the horizontal focusing will be lower than expected if the average field index $k_D$ of the defocusing magnet exceeds that of the focusing magnet.

\subsection{Monotonic behavior of the batatron wave numbers}

\subsubsection{Property}
In this section, we establish the following property for small scaling imperfections: \\ \\ \textit{In presence of scaling imperfections, the number of betatron oscillations per turn increases (resp. decreases) with the energy if $\kappa >0$ (resp. $\kappa <0$) where $\kappa=k_D-k_F$. Besides, the variation of the tune squared is, to the first order, proportional to $|\kappa|$, i.e. $\Delta(\nu_{x,y}^{2}) \approx a_{x,y} |\kappa|$.} \\  \\
From what preceded, it can be predicted that the tune variations are imposed by the AG term in the horizontal plane and the edge focusing as well as the AG term in the vertical plane. The average field index is almost insensitive to the azimuthal variations of $k(R,\theta)$. Let's start from the vertical plane by calculating the expression of the magnetic flutter. From \mbox{Eq. (\ref{zg_model}),} it results that:
\begin{eqnarray}
\mean{B^2} = &{B_{F0}}^2& \left( \dfrac{R}{R_0} \right)^{2 k_F} \nonumber \\
 &\times & \mean{\left[F_F(\theta) + A \left(\dfrac{R}{R_0}\right)^{\kappa} F_D(\theta) \right]^2}
\end{eqnarray}
where $A=a B_{D0}/B_{F0}>0$. Also,
\begin{eqnarray}
{\mean{B}}^2 = &{B_{F0}}^2& \left( \dfrac{R}{R_0} \right)^{2 k_F} \nonumber \\
 &\times & \mean{F_F(\theta) + A \left(\dfrac{R}{R_0}\right)^{\kappa} F_D(\theta)}^2
\end{eqnarray}
Thus, given that $F_F(\theta) F_D(\theta) =0$, and setting $B_{F0}$ and $B_{D0}$ such that \mbox{$\mean{F_F(\theta)}=-\mean{F_D(\theta)}=1$}, one obtains \mbox{$0<A<(R_0/R)^{\kappa}$} and
\begin{eqnarray}
\mathcal{F}^2 &=& \dfrac{\mean{B^2}}{\mean{B}^2}-1  \nonumber \\
&=& \dfrac{\mean{F_F^2(\theta)} + A^2 \mean{F_D^2(\theta)} \left(\dfrac{R}{R_0}\right)^{2\kappa}}{{ \left[1 - A \left(\dfrac{R}{R_0}\right)^{\kappa}\right]}^2}-1  \label{Eq:F2} \\ \nonumber
\end{eqnarray}
which is an increasing (resp. decreasing) function of the energy if $\kappa > 0$ (resp. $\kappa < 0$). Thus (when the AG is neglected) the vertical tune is an increasing (resp. decreasing) function of the energy if $\kappa>0$ (resp $\kappa<0$). 
Now, the magnetic flutter excursion writes
\begin{widetext}
\begin{eqnarray}
|\mathcal{F}^2(max) - \mathcal{F}^2(min)| &=& \left| \dfrac{\mean{F_F^2(\theta)}+A^2 \mean{F_D^2(\theta)} \left(\dfrac{R}{R_0}\right)^{2\kappa}}{{ \left[1 - A \left(\dfrac{R}{R_0}\right)^{\kappa}\right]}^2} - \dfrac{\mean{F_F^2(\theta)}+A^2 \mean{F_D^2(\theta)}}{{ \left[1 - A \right]}^2} \right| \\ \nonumber \\
&\approx &  \dfrac{2 \left[A \mean{F_D ^2(\theta)} + \mean{F_F^2(\theta)}\right]}{ \left(1-A  \right)^3} A  \dfrac{\Delta R}{R_0} |\kappa | \propto |\kappa| \label{deltanuy}
\end{eqnarray}
\end{widetext}
which partially proves the property above in the vertical plane. 
Now, to establish the same result in the horizontal plane, it is important to evaluate the monotonic behavior of the scalloping angle with the radius. Noting that:
\begin{eqnarray}
F(R,\theta) = \dfrac{B(R,\theta)}{B_m(R)} = \dfrac{F_F(\theta) + A \left(\dfrac{R}{R_0}\right)^{\kappa} F_D(\theta)}{1- A \left(\dfrac{R}{R_0}\right)^{\kappa}}
\end{eqnarray}
Integrating the previous equation while assuming $R$ constant, one obtains the expression of $\phi$:
\begin{eqnarray}
\phi(\theta) &\approx & \int_0^{\theta} \left[F(R,\theta)-1 \right] d\theta \nonumber \\
             &=& \dfrac{\widetilde{F_F}(\theta) + A \left(\dfrac{R}{R_0}\right)^{\kappa} \widetilde{F_D}(\theta)}{1- A \left(\dfrac{R}{R_0}\right)^{\kappa}}
\end{eqnarray}
yielding
\begin{eqnarray}
\mean{\phi^2} \approx \dfrac{\mean{\widetilde{F_F}(\theta)^2} + A^2 \left(\dfrac{R}{R_0}\right)^{2\kappa} \mean{\widetilde{F_D}(\theta)^2}}{\left[ 1- A \left(\dfrac{R}{R_0}\right)^{\kappa} \right]^2 }
\end{eqnarray}
which is in the same form given in Eq. (\ref{Eq:F2}) above. Thus $\mean{\phi^2}$ exhibits the same behavior as the magnetic flutter vis-\`a-vis field imperfections of the type described. The AG which is predominantly proportional to $\mean{\phi^2}$ follows as well. Qualitatively, the property established above states that the impact of reducing the average field index of the defocusing magnet in imperfect scaling FFA concept is to reduce the AG contribution as well as the magnetic flutter  impact on the transverse tunes. Stated differently, this is equivalent to reducing the FD ratio of the DFD triplet which decreases the focusing in both planes. 

Nevertheless, it is important to highlight that the validity of the above result fails when the magnet contains an overlapping region of the field of the $F$ and $D$ magnet. However, the above calculation can be extended to take this effect into account. \\
Equation (\ref{deltanuy}) shows that reducing the FD ratio helps reduce the tune variations. This is expected since, in our model, the D magnet is the source of the field defect. Furthermore, increasing the alternation of the gradient increases the sensitivity of the tunes to the field imperfections via the second order moments of the flutter functions, i.e. $\mean{F_D^2(\theta)}$ and $\mean{F_F^2(\theta)}$. In addition, it is shown that the effect of the scaling imperfections on the tune variations grows linearly with the radial excursion of the orbits in both horizontal and vertical planes. This shows the advantages of having an FDF triplet configuration (rather than a DFD one) with much larger field index, since then the orbit excursion can be reduced by a factor of 5 or more \cite{2nsstabi} leading to much lower tune variations due to field errors of the type described above.

\subsection{Tracking simulations}
Now, we demonstrate that our findings with the previous analysis are reinforced by numerical simulations and provide an application example which is the KURNS \mbox{150 MeV} scaling FFA. \\
From the hard edge model and the smooth approximation, it was found that the tune is sensitive to the average field index $k_F$ and $k_D$ of the $F$ and $D$ magnet respectively. In other words, breaking the scaling law, although a major source of imperfection in scaling FFA, can also be utilized in order to control the tune path in FFA. In order to quantify the source of imperfection, we introduce two new quantities in the calculation: the average value of the tunes $\nu _{x}^m = \mean{ \nu _x }$ and $\nu _{y}^m = \mean{\nu _y }$ over the range of energies to quantify the average focusing strength of the applied forces on the beam, and the rms value of the tunes  $\nu _x^{rms} = \sigma_{\nu_x}$ and $\nu _y^{rms} =\sigma_{\nu_y}$ to quantify the scaling imperfections in terms of tune variations. One could instead use the $ |max-min| $ value of the tunes to account for the oscillations. However, the rms quantities have the merit to be average quantities, thus more appealing to use in order to obtain smooth variations of the described quantities.

\subsubsection{Benchmarking work}
\begin{figure}[!h]
\centering
\includegraphics[height=8cm]{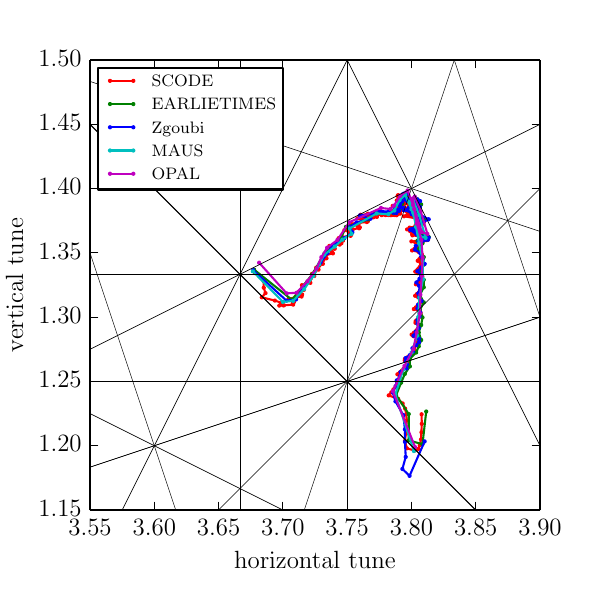}
\caption{Betatron tunes from 11 to 139 MeV (left to right) calculated with several codes \cite{benchm}. The Zgoubi model is in good agreement with the others. The solid lines, in black, show the resonance lines up to the $4^{th}$ order.}
\label{benchm_tunes}
\end{figure}

Following the FFAG14 workshop held at BNL\cite{FFAG14}, a simulation campaign was established to benchmark several simulation codes. The main objective is to provide reliable modelling tools for FFA type of accelerators and to better explain the results of the experiments at the KURNS 150 MeV scaling FFA \cite{suzie_charac}. There exists several simulation codes for FFAs \cite{smirnov}. However, only a few were part of this benchmarking analysis so far. For instance, the cyclops code which is probably the best known tool for beam dynamics simulations in cyclotrons can also be used for the simulation of FFA yielding results in excellent agreement with the others \cite{craddock1}. Nevertheless, the aim of the present paper is not to compare codes. \\
The first benchmarking test was carried out for the calculation of the betatron tune as a function of the momentum and shows excellent agreement between the different codes (that were part of this campaign) as shown in \mbox{fig. \ref{benchm_tunes}.}
The setup of the benchmarking model as well as the details of the simulation can be found in \cite{benchm}.

\subsubsection{Zgoubi tracking model}
\begin{figure}[!h]
\centering
\includegraphics[height=7cm]{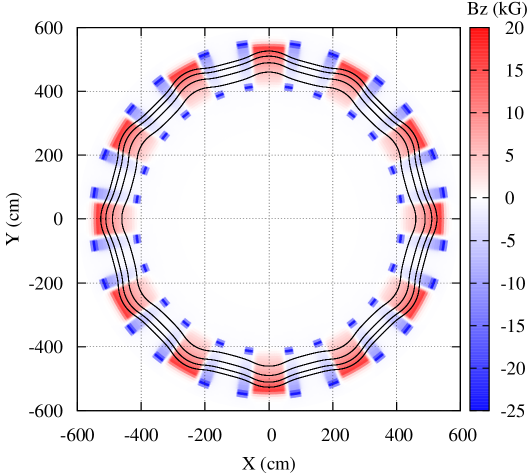}
\caption{ Example of several closed orbits for a scaling FFA. }
\label{closed_orbits}
\end{figure}

Based on the successful benchmarking test, we carry out parametric studies based on the zgoubi tracking code. The zgoubi code solves the non-linear equation of motion using truncated Taylor expansions of the field and its derivatives up to the $5^{th}$ order. Thus, it is more accurate than the linear approach. Given that the energy gain per turn is small in scaling FFAs (typically 2 keV per turn in the KURNS machine), one can reasonably assume that the accelerated orbit trajectory for any given energy is quasi the same as the closed orbit trajectory. Thus, the procedure employed for the calculation of the betatron wave number is based on the closed orbits formalism described below: \\
First, one generates a median plane field map for a given $(a,k_F,k_D)$ as shown in fig. \ref{field_co}. The field fall-off at the end of the magnets is obtained by using an Enge-type fringe field model \cite{enge}. Extrapolation off the median plane is then achieved by means of Taylor series: for that, the median plane symmetry is assumed and the Maxwell equations are accommodated. This yields results in excellent agreement with the 3D field map calculation. \\
Second, search for NCO closed orbits between injection and extraction using the built-in fitting routines in zgoubi. NCO was chosen to be 30 in order to have good statistics and ensure the convergence of the calculated quantities. A typical example of 4 closed orbits search is illustrated in fig. \ref{closed_orbits}.
 \\
Lastly, for each closed orbit, the betatron wave number is calculated in both planes.

\begin{figure}[!h]
\centering
\includegraphics[height=7cm]{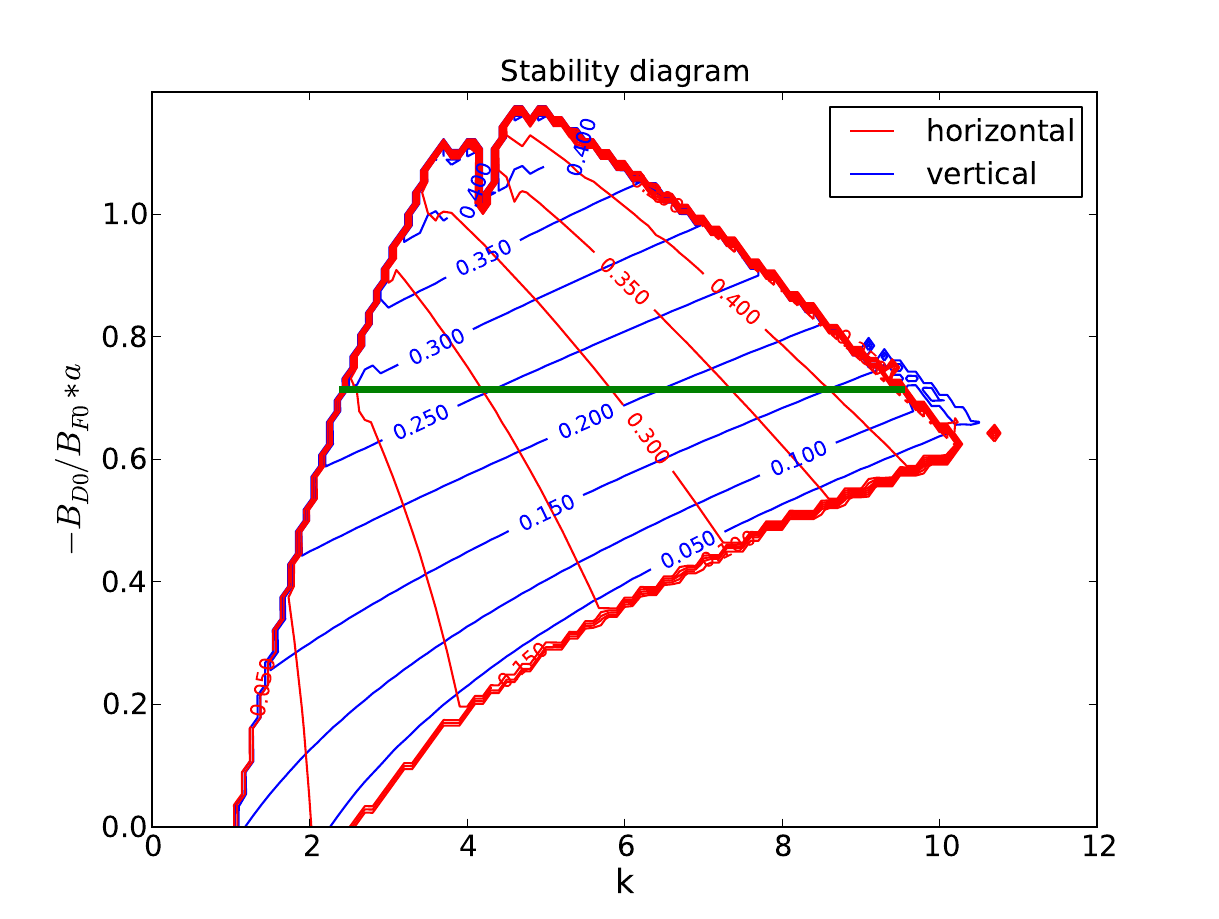}
\caption{ Stability diagram as a function of the FD ratio and the average field index $k$: the green line shows the value of the FD ratio that we choose for the study that follows. The number along with the red and blue lines are the horizontal and vertical
cell tunes, respectively. Note that the region on the left side is limited by the physical size of the magnets, which is limited to $R_{max}$ = 10 m.}
\label{FDra}
\end{figure}

Fig. \ref{FDra} shows the stability diagram obtained by varying the average field index of the magnets ($\kappa = 0$) as well as their FD ratio, therefore the scale factor $a$ in \mbox{Eq. (\ref{zg_model}).}  
One can observe that, on the top and bottom right, the stability limits are set by the horizontal and vertical cell tunes, respectively. On the left side, the physical size of the magnets (here we choose a radial excursion limited to 10 m) determines the boundary limits.

\begin{figure}[!h]
\centering
\includegraphics[height=7cm]{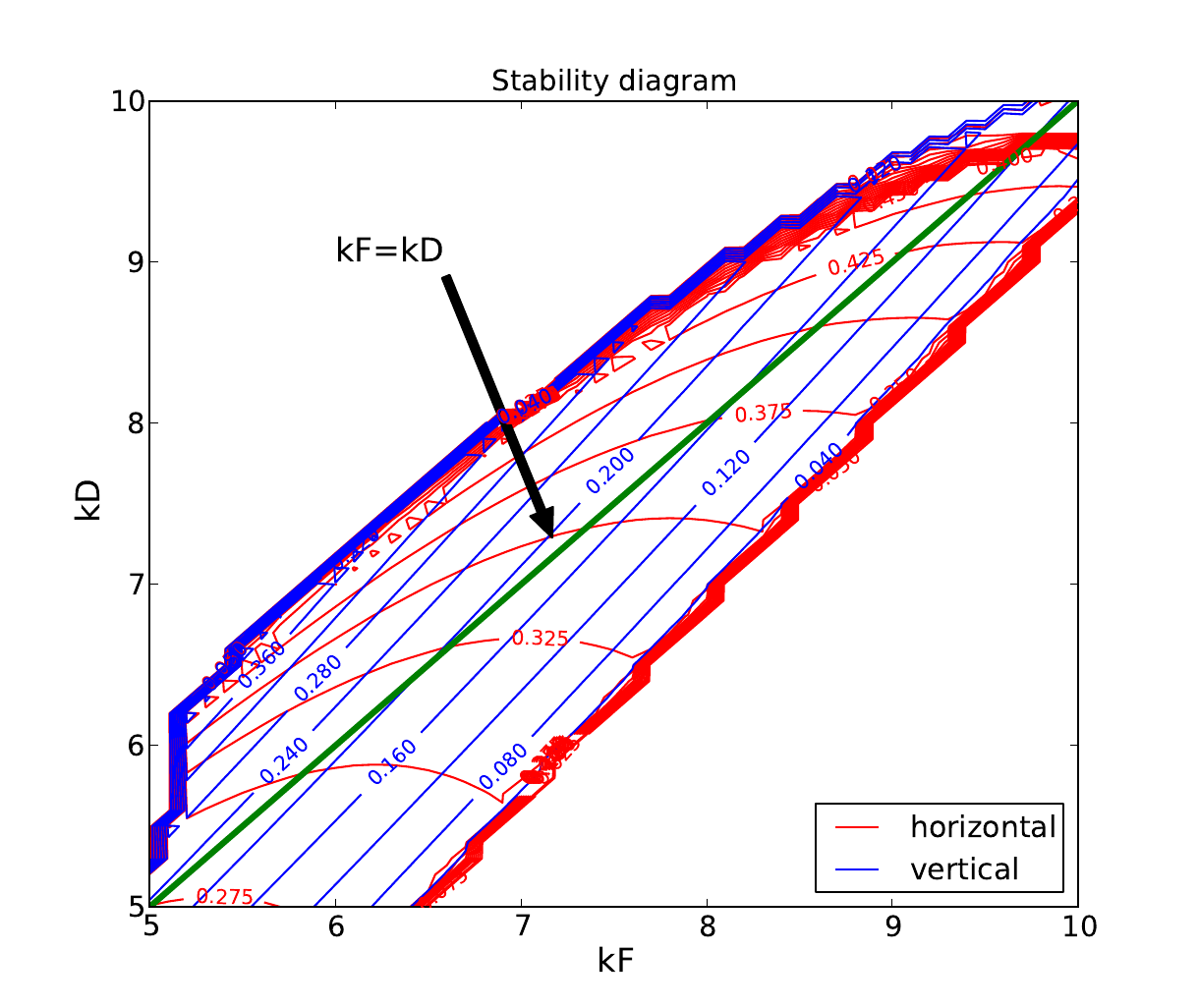}
\caption{ Stability diagram of the average cell tune as a function of the average field index $k_F$ and $k_D$: the green line corresponds to the case of a scaling FFA with no field imperfections, i.e. $\kappa=0$.}
\label{kFkD}
\end{figure}

Now, we choose to focus our analysis on the average field index of the magnets. For that, we fix the FD ratio. We choose $a=1$ which corresponds to the green line in fig. \ref{FDra}. A scan on $k_F$ and $k_D$ provides the stability diagram of the DFD triplet in the transverse plane (see fig. \ref{kFkD}). Qualitatively, it shows that, in the case where $k_F = k_D = k$, the average cell tune exhibits the expected behavior predicted by the Symon formula i.e. Eq. (\ref{eq:symon_tunes}): increasing $k$ increases the horizontal tune and decreases the vertical one. One can also observe that for large $k$ values, the stability diagram shrinks, thus any design imperfection will make the orbits quickly unstable. This is explained in the following way:
on the right side of the stability diagram, i.e. when $\kappa<0$, the stability limit is set by the condition that $\nu_y^2$ is to remain positive (Floquet resonance), given that the tunes decrease with the energy. \\
On the left side of the stability diagram, i.e. when $\kappa>0$, the stability limit is set by the radial $\pi$-mode stop-band resonance, given that the tunes increase with the energy. \\
Note that a second stability island exists for larger \mbox{$k$ values \cite{2nsstabi}.} However, we restrict our analysis to a KURNS type FFA for which the design value of $k \approx 7.6$.

\begin{figure}[!h]
\centering
\includegraphics[height=7cm]{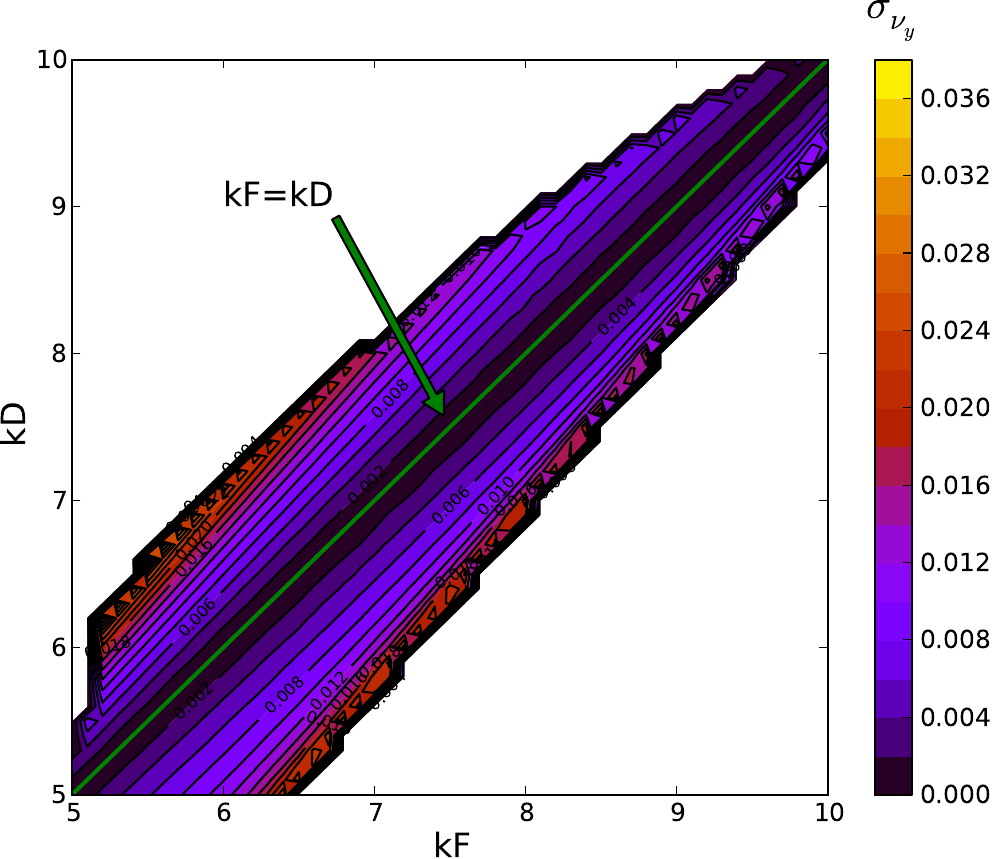}
\caption{ Stability diagram of the rms vertical cell tune as a function of the average field index $k_F$ and $k_D$: the green line corresponds to the case of a scaling FFA with no field imperfections, i.e. $\kappa=0$. Note that a similar result is obtained for the horizontal plane.}
\label{rmsy}
\end{figure}
Now, calculating the RMS tune variations shows that the latter exhibit the expected behaviour in the vicinity of the line $k_F = k_D$ where they become negligible. This is shown in fig. \ref{rmsy}. When field imperfections such that \mbox{$\kappa \neq 0$} are introduced, one can observe that the tune variations increase with $|\kappa|$ as demonstrated earlier. 

\begin{figure}[!h]
\centering
\includegraphics[height=6cm]{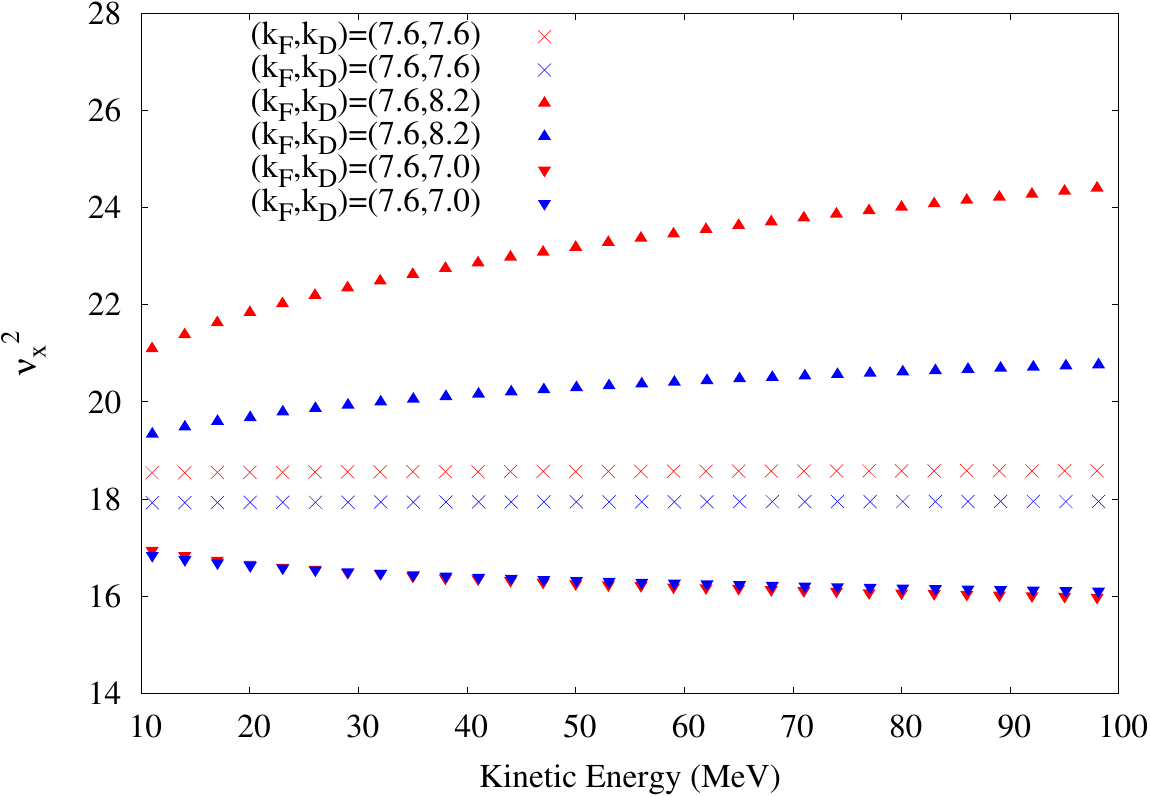}
\caption{ Example of tune calculation using the  tracking code zgoubi (red) and comparison with the 3rd order analytical formula \ref{Eq:tune_BKM} (blue). For the case $(k_F,k_D)=(7.6,8.2)$, the large discrepancy observed ($ \lesssim 13 \%$) is due to the fact that the beam is located nearby the boundary of stability. }
\label{tunex_bogo1}
\end{figure}

\begin{figure}[!h]
\centering
\includegraphics[height=6cm]{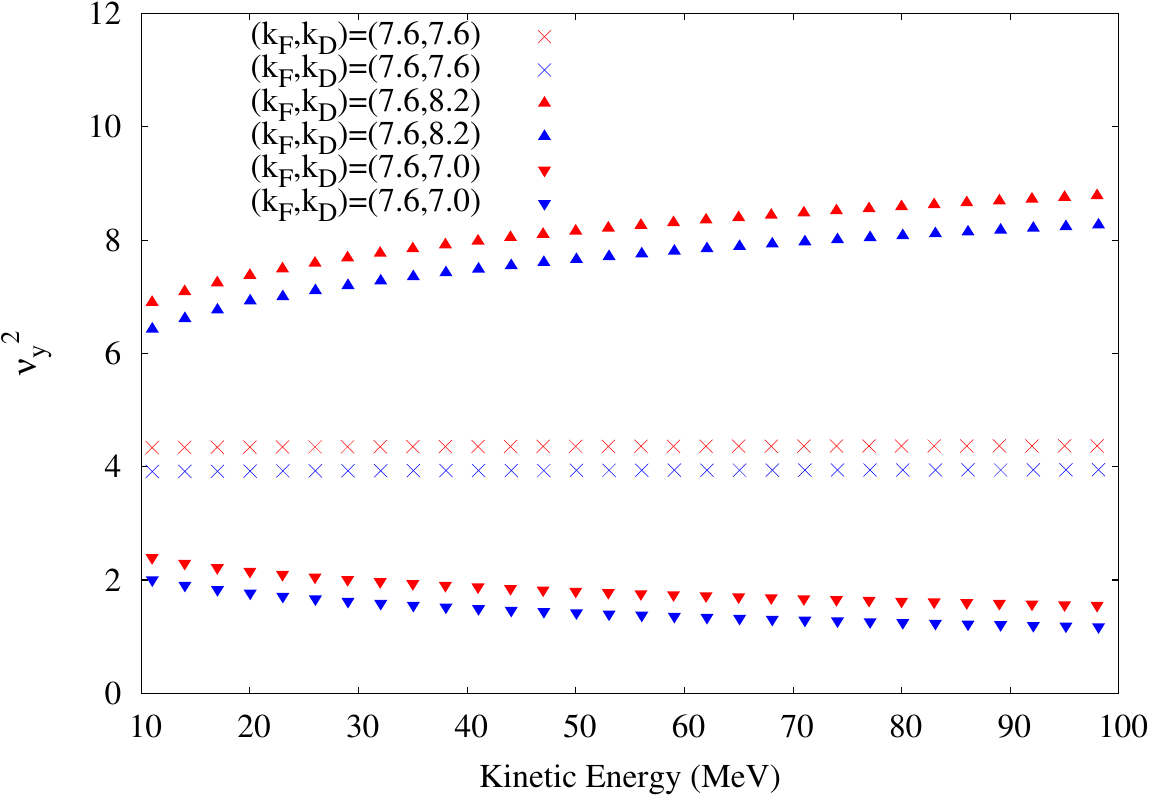}
\caption{ Example of tune calculation using the  tracking code zgoubi (red) and comparison with the 3rd order analytical formula \ref{Eq:tune_BKM} (blue). To the third order, good agreement \mbox{($< 10 \%$)} is obtained in all cases considered. }
\label{tuney_bogo1}
\end{figure}

Based on all the above, we compare the tracking results with those obtained from the analytical formula established in the previous section. This is shown in figs. \ref{tunex_bogo1} and \ref{tuney_bogo1} for the horizontal and vertical plane, respectively. The red points are the simulation results while the blue points represent the analytical formula: for $\kappa > 0$ (upward-pointing triangle), the tunes increase with the energy in both planes, while for $\kappa < 0$ (down-pointing triangle), the tunes exhibit the opposite behaviour. This confirms the findings of the previous section.  

\begin{figure}[!h]
\centering
\includegraphics[height=6cm]{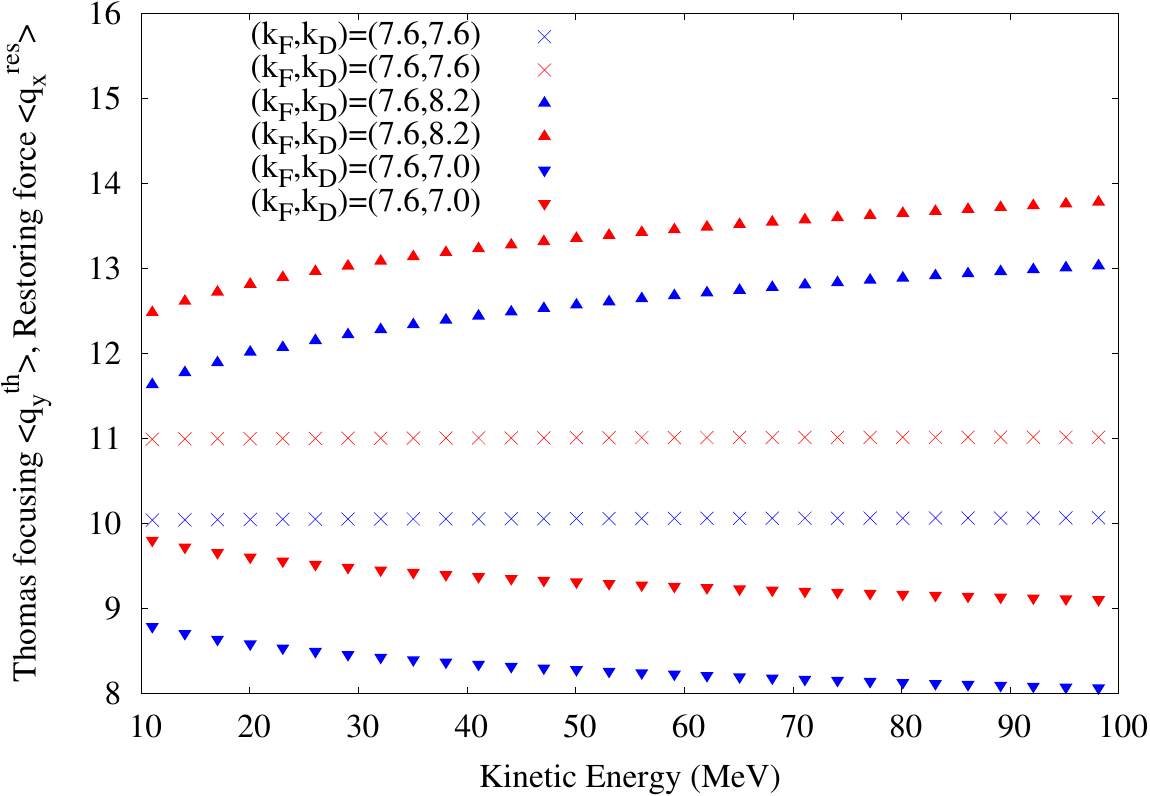}
\caption{The horizontal restoring force of the closed orbits (in red) as well as the Thomas focusing (in blue) as a function of the kinetic energy.}
\label{flutterr}
\end{figure}

The main finding of the smooth approximation is that scaling imperfections produce an orbit distortion that manifests through a radial dependence of the scalloping angle of the orbits as well as the magnetic flutter. The tracking simulations confirm our findings: as shown in fig. \ref{flutterr}, the Thomas focusing explains the monotonic behavior of the vertical tune as a function of the energy for various $\kappa$ values. Nevertheless, in the horizontal plane, the horizontal restoring force and the Thomas defocusing effect act in opposition such that Eq. (\ref{Eq:qxres_qyth}) holds. Thus, it is the AG that explains the monotonic behavior of the tunes in the horizontal plane (the average field index $k$ of the orbits changes insignificantly).
  
\subsection{Application to the KURNS 150 MeV scaling FFA }
\begin{figure}[!h]
\centering
\includegraphics[height=5.5cm]{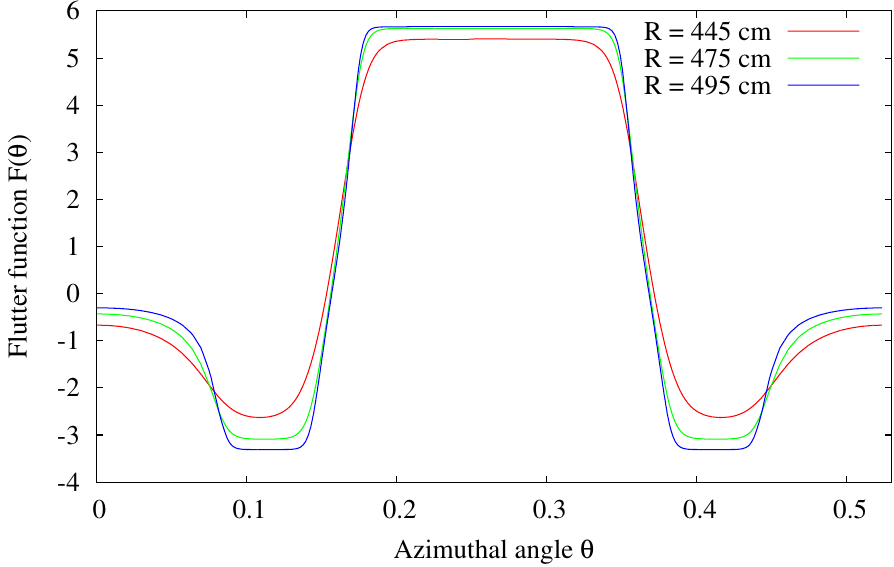}
\caption{ Flutter function at different radii of the KURNS 150 MeV FFA.}
\label{flutter_KURRI}
\end{figure}

\begin{figure}[!h]
\centering
\includegraphics[height=5.5cm]{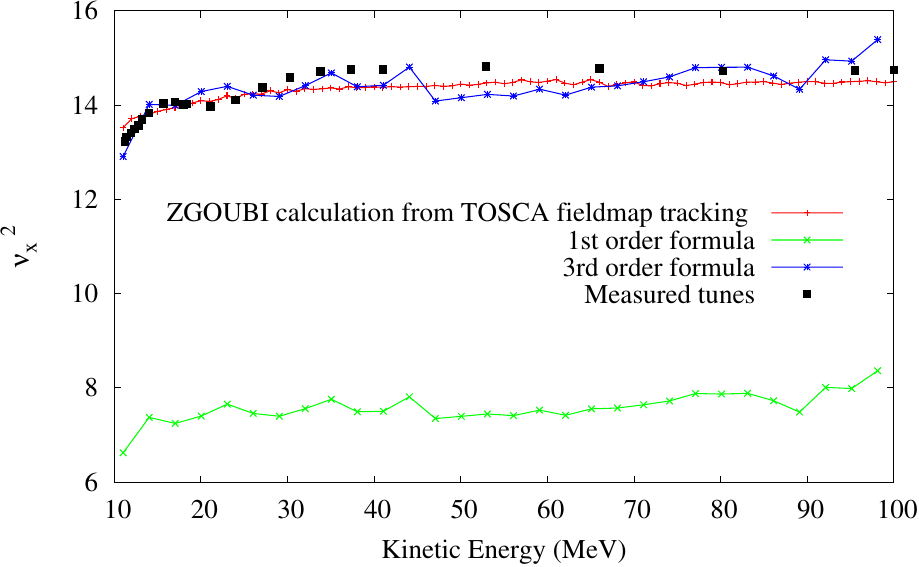}
\caption{ Tune calculation in the horizontal plane of the KURNS 150 MeV FFA ring and comparison with the analytical formula as well as the measurement.}
\label{tunex_compar}
\end{figure}

\begin{figure}[!h]
\centering
\includegraphics[height=5.5cm]{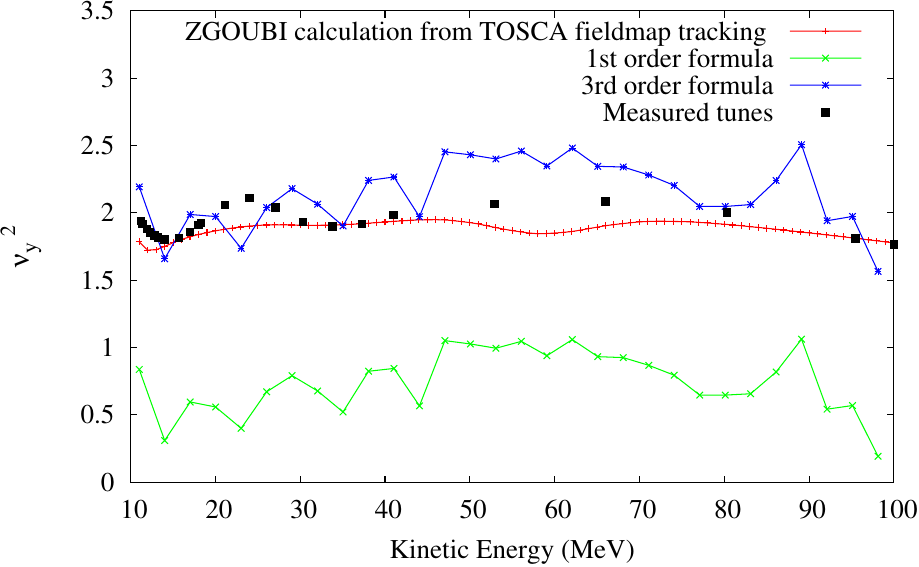}
\caption{Tune calculation in the vertical plane of the KURNS 150 MeV FFA ring and comparison with the analytical formula as well as the measurement. Note the oscillatory behavior of the 1st and 3rd order approximation which is due to the fact that the flutter function $F$ is calculated only at 3 different locations in the ring.}
\label{tuney_compar}
\end{figure}

We will benchmark the analytical formula against the simulated values obtained from particle tracking using 3D field maps and compare with the measurement. \\
The tracking results provide the closed orbits and therefore the scalloping angle $\phi$ for various particle energies which are then exploited to calculate the horizontal and vertical tunes by applying the 1st and 3rd order approximation given by Eq. (\ref{Eq:tune_BKM}). The difference between the two approximations lies in the fact that the AG effect is accounted for by the higher order terms which represents approximately half the focusing in both planes. \\
Each magnet is characterized by an average field index which is assumed to be constant, $k_F=7.6$ and $k_D=9.0$. In addition, the flutter function $F$ changes considerably with the radius. However, the latter is re-calculated only at 3 different radii as shwon in fig. \ref{flutter_KURRI}. \\
Comparison between the different results are summarized in figs. \ref{tunex_compar} and \ref{tuney_compar}. To begin with, the 1st order formula is less in agreement with the tracking results. The main reason is the missing focusing which results from the AG effect. In addition, it is important to note that the monotonic behavior of the tunes is a consequence of several localized imperfections that our simplified model does not take into account. In particular, due to the variations of the flutter function with the radius, it may be interesting to calculate the spiral focusing term $q_y^{spi}$ to determine its impact on the beam dynamics. However, this is not what we seek in this model which aims at simplifying the conception of the imperfections at the KURNS FFA and yields satisfactory results so far. 

\section{Correction scheme and advanced FFA concept}
\label{section4}
Practically, it is difficult to correct the orbit and optics distortion in fixed field accelerators for the entire momentum range since the beam moves outward radially during acceleration. Therefore, a dedicated correction system should be implemented along the radius of the magnet to produce the desired field profile. From the point of view of cyclotrons, this consists in the implementation of trim coils to correct the isochronism, i.e. the revolution time of the orbits. From the point of view of a scaling FFA, the main target is to fix the betatron wave number in both planes which will allow to avoid the crossing of transverse resonances and maximize the overall beam transmission from injection to extraction. \\
From what preceded, one obtained general rules to explain the monotonic behaviour of the tunes as a function of the energy as well as the amplitude of its variations. This is a crucial result if one aims to reduce the tune excursion.
One major outcome of this study is that gradient errors in the FFA magnet yield a non-scaling of the orbits and lead to a change of the average as well as the alternating gradient focusing forces applied on the beam. This means that fixing the field defect of the FFA magnet by aiming to produce a constant average field index $k$ (by considering the average field over the entire circumference) is not sufficient since the azimuthal variations of this quantity yield a non-scaling of the orbits. In what follows, we present a novel scheme to correct the field errors in FFA that relaxes the constraint of having the scaling law valid everywhere in the ring.
\subsection{Alternating scaling imperfections}
In the context of the present study and for the sake of simplicity, the field defect is due to one of the magnets, either the focusing (F) or the defocusing (D) one such that $\kappa \neq 0$. Without any loss of generality, we assume that the D magnet is the source of imperfection. In order to minimize the tune variations, one way is to reduce the FD ratio of the DFD triplet. This can be achieved by reducing its excitation current or by sliding it to outer radii so that the average field encountered by the particle at any radius is lower. However, this approach leads to the loss of focusing in both planes as shown earlier (see fig. \ref{FDra}). Another interesting but not yet evaluated approach consists in misaligning vertically some of the magnets in order to modify the shape of the closed orbits. However, this is not considered in the present paper since one neglected the changes in the vertical direction in Eq. (\ref{Eq:ds_incr}). Based on the property established earlier, the tune variation with the energy exhibits antagonistic behavior based on the sign \mbox{of $\kappa$.} Therefore, the idea of the following correction scheme is to introduce a perturbation of the field every $P$ sectors in order to counteract the already existing imperfections. For instance, if we choose $P=2$, then a 12-fold symmetry machine is replaced by a 6-fold symmetry in the following way: let's note $D_i$ (resp $F_i$) the Defocusing (resp Focusing) magnet with scaling factor $k_{Di}$ (resp $k_{Fi}$). The original design \mbox{$12 \times (D_0F_0D_0)$} is replaced by \mbox{$6\times (D_0F_0D_0D_1F_0D_1)$} in the following way: if \mbox{$k_{D0}>k_{F0}$} then \mbox{$k_{D1}<k_{F0}$} and vice-versa.
Thus, instead of aiming to fix the design imperfections by correcting the field profile to match with the ideal one for every magnet and make the orbits scale at every azimuthal position, one can fix the scaling of the orbits on an average sense by creating alternation of the difference of the average field index of the magnets. This has a major advantage of reducing the cost of the correction system since then only 12 D-magnets (out of 24 magnets) will require trimming coils to be implemented. The number can be further reduced if increasing $P$. This is the cornerstone of the fixed tune non-linear non-scaling radial sector FFA concept that is discussed in detail in what follows.

\subsection{Fixed tune non-scaling FFA}
\begin{figure}[!h]
\centering
\includegraphics[height=8cm]{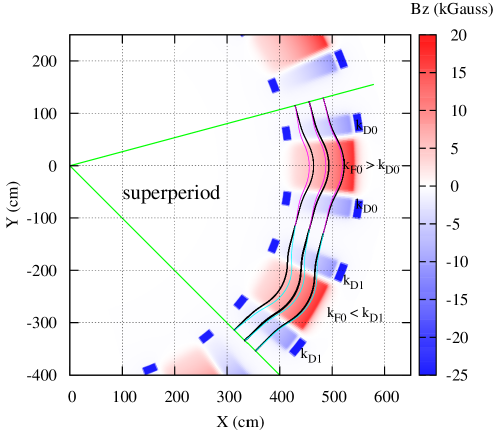}
\caption{Closed orbits in a scaling FFA (black) and in a fixed tune non-scaling FFA with alternating $\kappa$ (pink and light blue). For the sake of clarity, the distance between the closed orbits of the scaling and the non-scaling FFA is amplified.}
\label{orbit_advan_conc}
\end{figure}

\begin{figure}[!h]
\centering
\includegraphics[height=8cm]{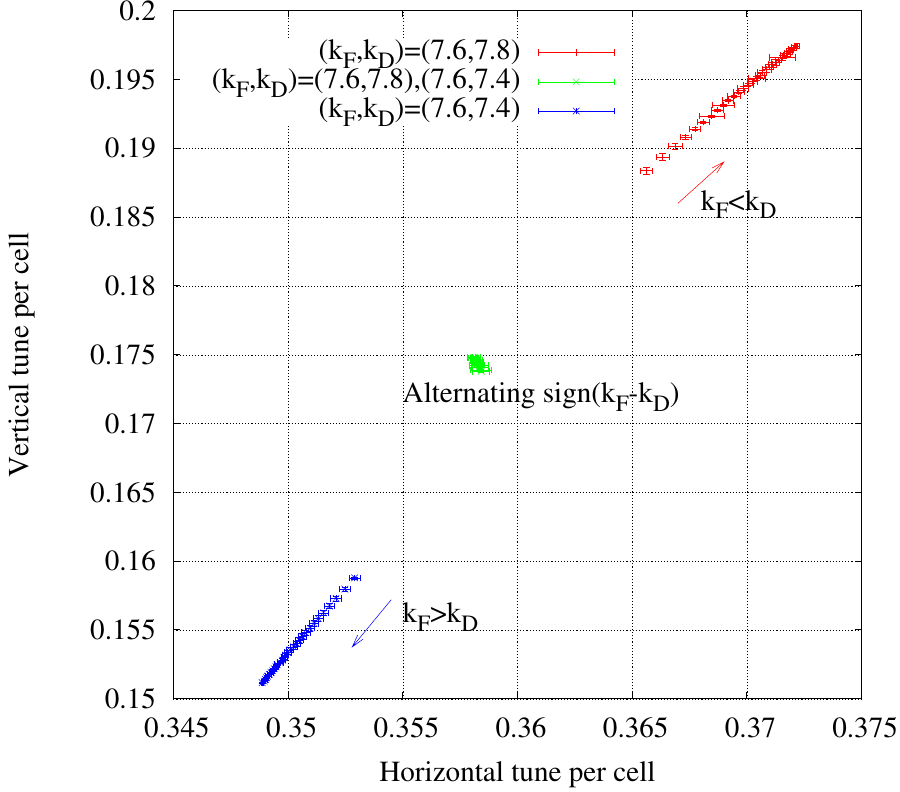}
\caption{Tune variations as a function of the energy before and after correction: the corrected scheme is shown in green where $(k_{F0},k_{D0})=(7.6,7.8)$ and $(k_{F1},k_{D1})=(7.6,7.4)$. These results are obtained from multi-particle tracking assuming a Gaussian distribution of the particles in the transverse plane. The errorbars represent the overall tune variation from the distribution.}
\label{correcttunxy}
\end{figure}

\begin{figure}[!h]
\centering
\includegraphics[height=6cm]{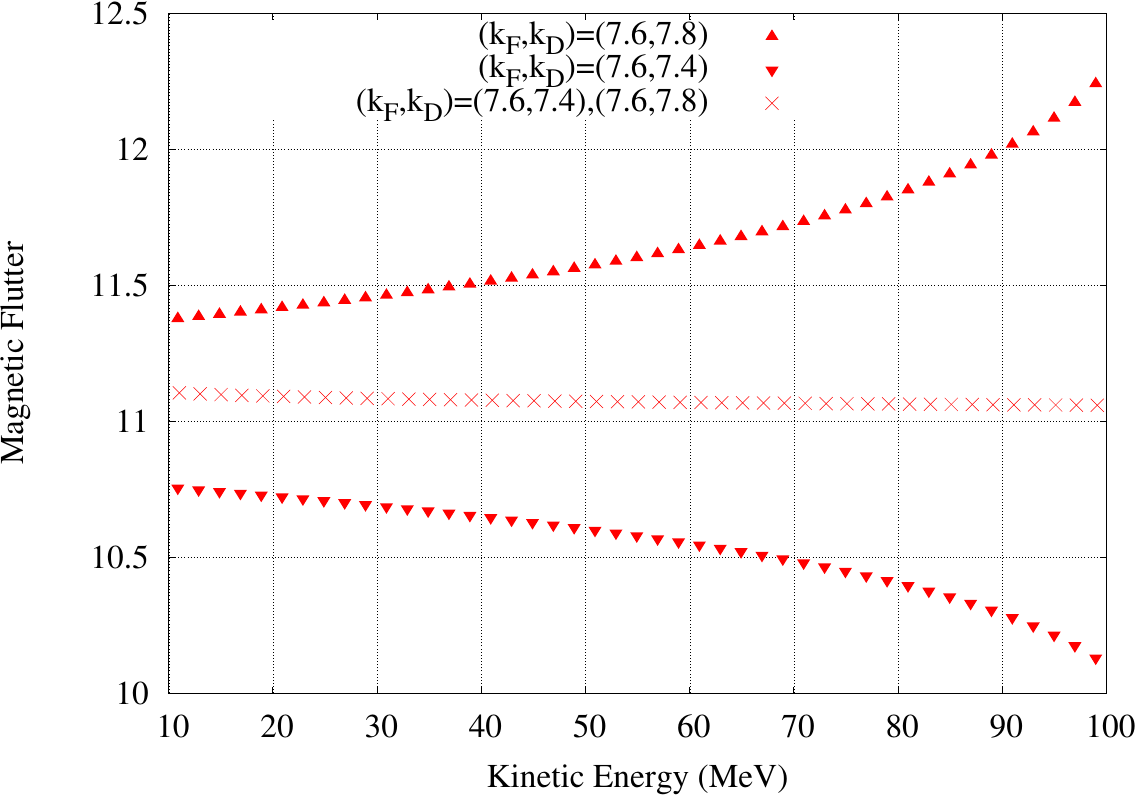}
\caption{Magnetic flutter in the advanced fixed tune FFA concept. }
\label{flutter_advan}
\end{figure}

In what follows we shall examine the characteristics of this concept which incorporates the alternation of the difference of the gradients scheme to FFA. The particle trajectories in the median plane are shown in fig. \ref{orbit_advan_conc} where one compares the orbits of a 12-fold symmetry scaling FFA $12\times(k_{F0},k_{F0})$ with those of a 6-fold symmetry concept \mbox{$6\times(k_{F0},k_{D0}),(k_{F0},k_{D1})$:} the orbit scalloping in the non-scaling case differs mainly in the dominant F-magnet. If $\kappa<0$, the radial field increase in the F-magnet is faster than in the D-magnet so that the equilibrium orbit (pink) in the F-magnet is at lower radii compared to the scaling case. The opposite effect occurs when $\kappa>0$ (light blue orbit).
As a consequence of the alternation of $\kappa$, the monotonic behavior of the phase advance per cell is alternating (increasing if $\kappa > 0$ and decreasing if $\kappa<0$). This is illustrated in fig. \ref{correcttunxy} where one can see that the combination of the two cells yields a fixed average tune per cell.
This results from the alternation of the monotonic behavior of the horizontal restoring force as well as the Thomas focusing and the AG effect which is due to the azimuthal change of the orbit scalloping angle $\phi$. The Thomas focusing term $\mean{q_y^{th}}$ within each cell as well as for their combination is shown in fig. \ref{flutter_advan} and is in agreement with the property established above. The same holds for all other quantities.

\subsection{Dynamic Acceptance}
\begin{figure}[!h]
\centering
\includegraphics[height=6cm]{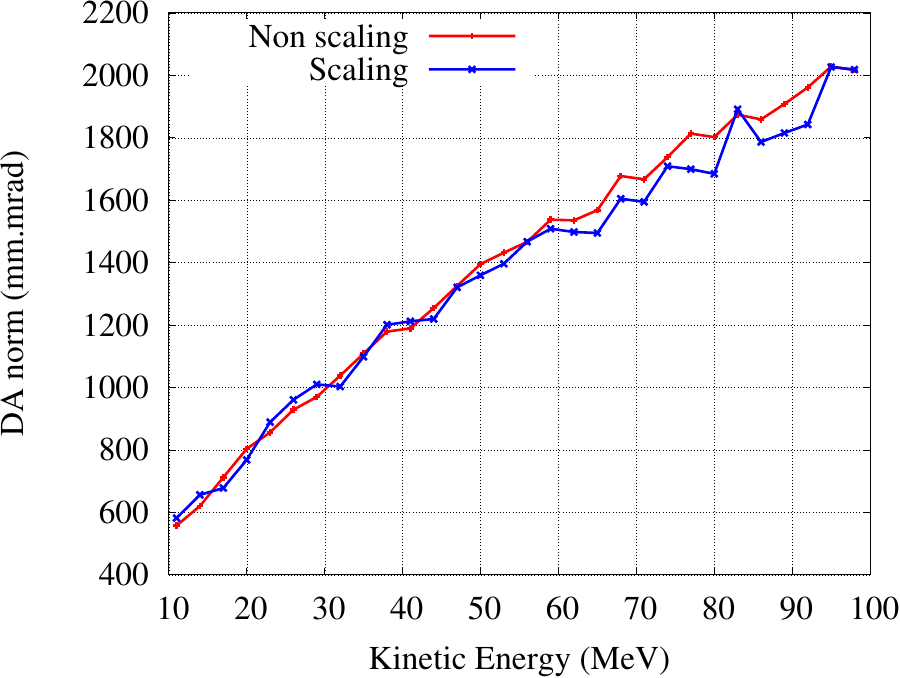}
\caption{Comparison of the DA of the scaling and the non-scaling FFA in the horizontal plane.}
\label{da_scal_non_scal}
\end{figure}

\begin{figure*}[htp]
  \centering
  \subfigure[Case of scaling FFA]{\includegraphics[scale=0.4]{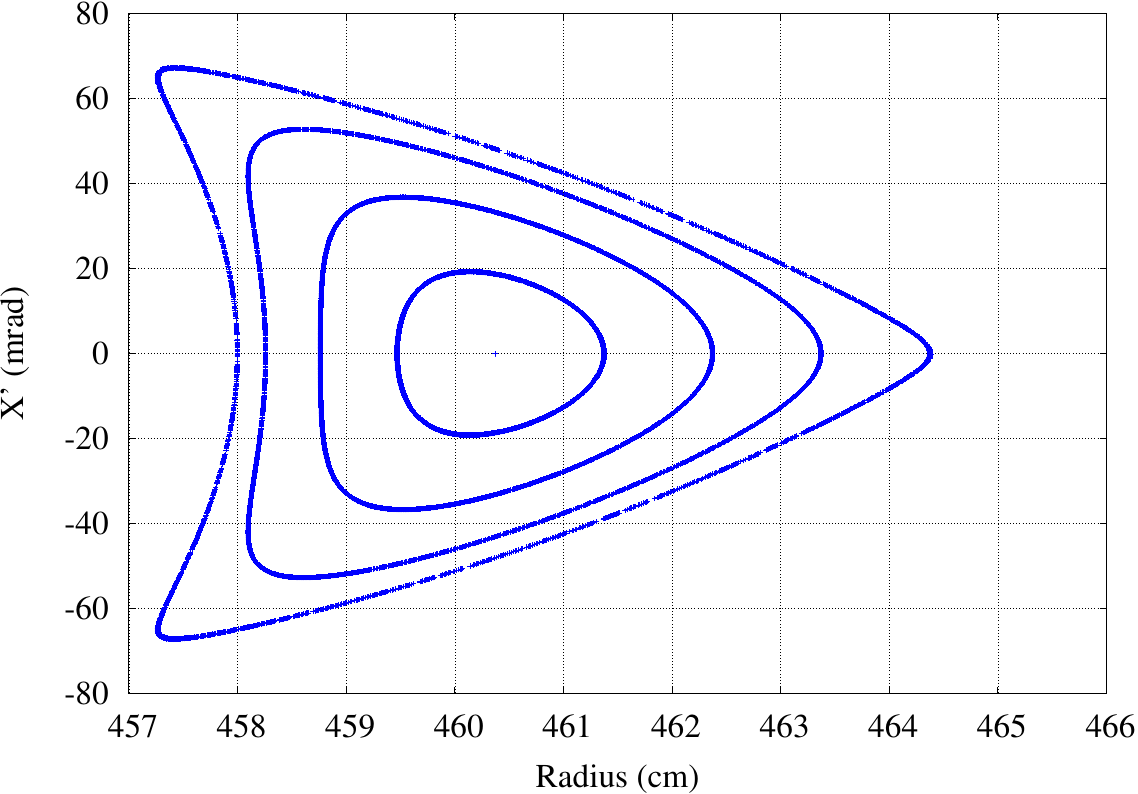}}\quad
  \subfigure[Case of non-scaling FFA: the trajectories in the $1^{st}$ cell are shown in red while the trajectories in the $2^{nd}$ cell are shown in green.]{\includegraphics[scale=0.4]{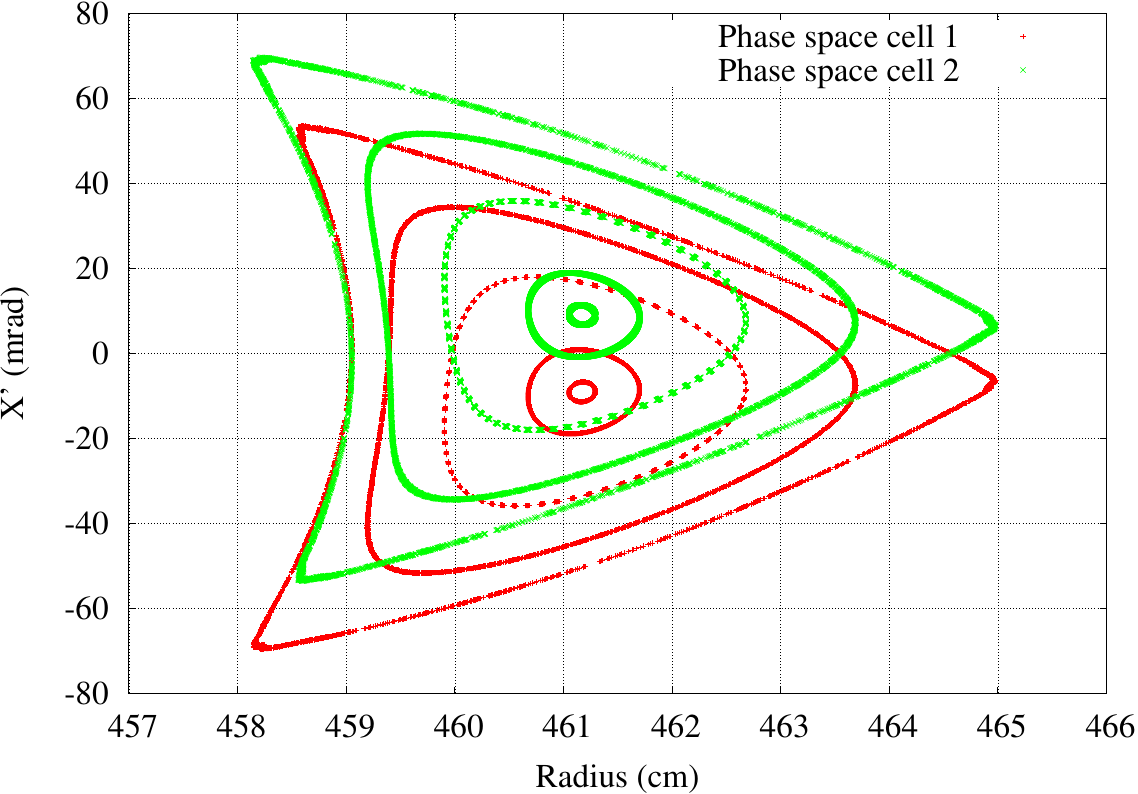}}
  \caption{Horizontal phase space trajectories at 100 MeV including the separatrix.}
  \label{phase_space}
\end{figure*}

Although strong non-linearities of the field are inherent to the scaling FFA, large Dynamic Acceptance (DA) is obtained with this concept. Therefore, one main question to answer is how the DA of the fixed tune non-scaling FFA compares to that of the scaling FFA. \\
In our analysis, the DA is defined as the maximum transverse invariant value that the beam can have without loss due to single particle dynamics effects. Particle tracking at fixed energy is employed for our analysis. A particle with original displacement from the closed orbit is defined as stable if it survives 1000 turns. Given that the vertical aperture in fixed field accelerators is the limiting factor due to the small gap size, we focus our analysis on the horizontal plane. However, in our simulations, it is noted that a vertical beam size up to 1 cm at injection can be transported without any losses and that the horizontal DA is insensitive to it. The main idea is to generate two lattices that have the same tunes in both planes. This is obtained by first generating a non-scaling fixed tune lattice with $\kappa \approx 0.3$, then matching its tunes by finely tuning the average field index as well as the FD ratio of the scaling lattice. This is achieved for a lattice with \mbox{$(\nu_x,\nu_y)=(4.43,2.16)$.}  \\
Comparison of the calculated DA in both cases shows \mbox{(fig. \ref{da_scal_non_scal})} that, for the same tunes, the horizontal beam acceptance is the same even though the orbits do not scale. This is only valid if resonance crossing is not occurring: in the non-scaling case discussed here, and for $P=2$, the resonance population is doubled in comparison with the scaling case. Thus, one can expect that the resonance crossing problem is more severe in the non-scaling case. This requires further investigation.
Comparison of the phase space trajectories between the two cases is finally shown in fig. \ref{phase_space}. The trajectories in both cells are symmetric with respect to $X'=0$.

\section{Conclusion}
In this paper, one analyzed the stability of the particle trajectories due to field errors. Several approaches to the problem were developed and analyzed. Comparison of the results showed that the first order approximation based on the smooth approximation is only sufficient for a lattice where the average field index is negligible. Relying on the non-linear approach based on tracking simulations alongside the analytical derivations based on the 3rd order approximation from the smooth approximation, a crucial result was to establish a relationship between the betatron wave number and the field defects. A key parameter to measure the amplitude of the defects is the $\kappa$-value defined as the difference of the average field index of the focusing and defocusing magnets. Furthermore, analysis of the stability diagram (fig. \ref{kFkD}) showed that the tolerance to scaling imperfections becomes lower when increasing the average field index of the magnets. Based on these results, a new scheme to remediate the variation of the betatron oscillations with the energy was proposed. The main idea consists in alternating the $\kappa$-values of the magnets, every two (or more) sectors. This leads to the new concept of the fixed tune non-scaling radial sector FFA that one developed in section \ref{section4}: in addition to the fact that this demonstrates that the conditions of scaling are non necessary to obtain a fixed tune FFA, the newly developed concept is easier to implement by means of trim coils that can be adjusted to find the condition of minimum tune excursion and avoid the crossing of harmful resonances. Analysis of the DA showed that the lattice with alternating $\kappa$-values has the same DA as the equivalent scaling FFA case.
Given that the alternating-$\kappa$ FFA reduces the number of super-periods in the accelerator, therefore doubling the resonance population, one can expect that the impact of the resonance crossing is more severe than the scaling FFA case. This needs further investigation.
\\ \\

\begin{acknowledgements}
The authors would like to express their special gratitude to the members of the international Kyoto University Research Reactor Institute collaboration. The research leading to these results has received funding from Brookhaven Science Associates, LLC under Contract No. DE-SC0012704 with the U.S. Department of Energy. 
\end{acknowledgements}

\appendix

\section{Second order approximation}
\label{appendix:b}
In this appendix, one establishes an improved approximation of the orbit scalloping angle. The main idea is to use the first order approximation given by Eq. (\ref{Eq:F_phi}) and use the method of successive approximations for the set of coupled equations (\ref{Eq:tan}),(\ref{Eq:F_phi_ex}) in order to determine an improved second order approximation of the scalloping angle of the closed orbit. In this analysis, one shall assume a radial sector FFA so that the flutter function does not evolve with the radius. \\
Again, under the assumption of small orbit scalloping, Eq. (\ref{Eq:tan}) can be integrated \mbox{($\tan(\phi) \approx \phi $)} and the radius of the closed orbit is given by:
\begin{eqnarray}
R(\theta) &\approx & R(0) \exp\left(- \int_0^{\theta} \phi(\theta) d\theta \right) \\
          &\approx & R(0) \exp\left(- \int_0^{\theta} \widetilde{F}(\theta) d\theta \right)  \label{Eq:Rapprox}
\end{eqnarray}
where $\widetilde{ }$ is the integrating operator defined by: \mbox{$\widetilde{F}(\theta)= \int_0^{\theta} \left(F-\mean{F}\right) d\theta$.}
Now, from Eq. (\ref{Eq:F_phi_ex}), the expression of the scalloping angle can be re-written in the general form (where one assumed \mbox{$\cos(\phi) \approx 1$}):
\begin{eqnarray}
1+\dot{\phi} \approx \dfrac{B_m(R(\theta)). R(\theta). F(\theta)}{\mean{B_m(R(\theta)). R(\theta). F(\theta)}}
\end{eqnarray}

Finally, assuming that the average magnetic field $B_m(R)$ obeys the $R^k$ law of a scaling FFA yields:
\begin{eqnarray}
1+\dot{\phi} &\approx & \dfrac{\exp\left[- (k+1)\int_0^{\theta} \widetilde{F}(\theta) d\theta \right]}{\mean{\exp\left[- (k+1)\int_0^{\theta} \widetilde{F}(\theta) d\theta \right] F(\theta)}} F(\theta) \nonumber \\
&=& h(\theta) F(\theta)
\end{eqnarray}
which shows that the scalloping angle is a function of the average field index of the magnet as well as the flutter function. Comparison of the second order approximation with the results of tracking is finally shown in fig. \ref{fig:flutter_vs_orbit} where one obtains a good agreement. To illustrate the impact of the average field index on the scalloping angle of the closed orbit, one assumes a lattice for which the flutter function is given by: $F(\theta)=1-3 \cos(12\theta)$ and plot the function $h(\theta)$ for various $k$ values. It results that the orbit scalloping decreases with increasing $k$ hence the edge focusing decreases as well as shown in section \ref{subsection:benchm}.
\begin{figure}
\centering 
\includegraphics*[width=7cm]{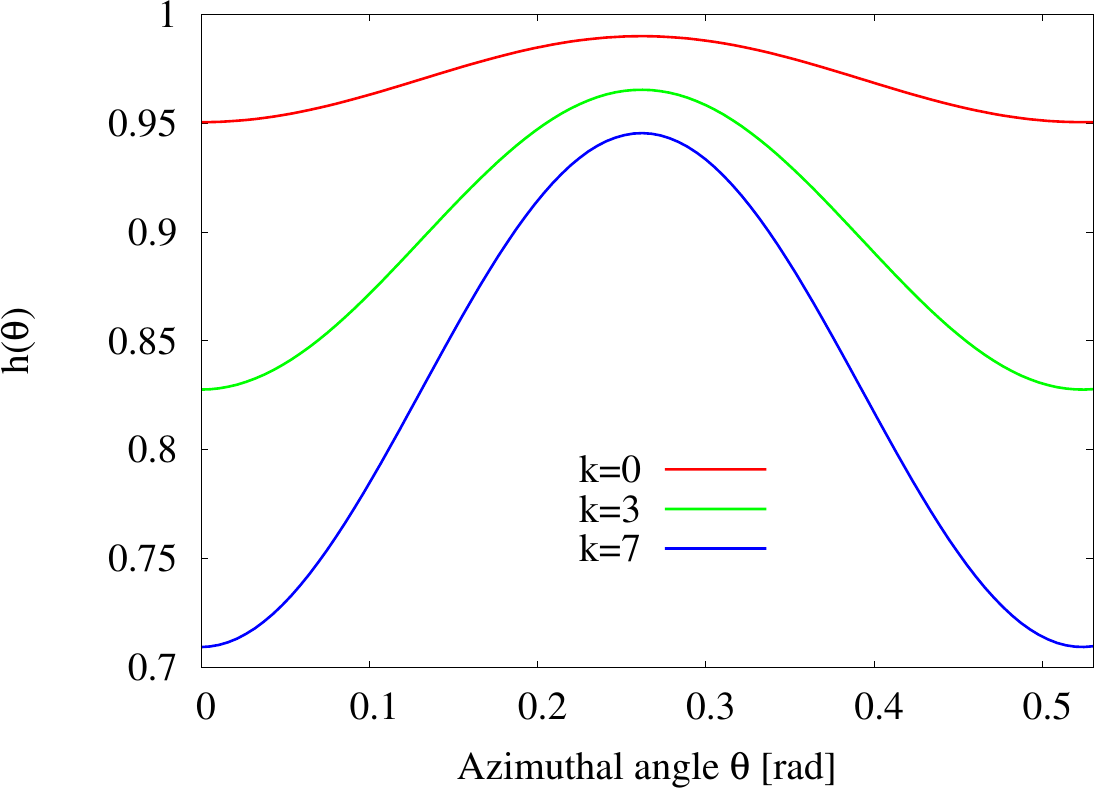}
\caption{Plot of the function $h(\theta)$ for various $k$ values.}
\label{fig:hfunction}
\end{figure}

\section{Spiral focusing}
\label{appendix:a}
Let's assume that the magnet boundaries are deformed from radial poles to spiral shaped poles. As an example, the flutter function of the median plane magnetic field for a spiral sector FFA can be written in the following way \cite{symon}:
\begin{eqnarray}
B(R, \theta) &=& B_0 \left(R/R_0\right)^k F(R,\theta) \\
F(R, \theta) &=& 1 + f \cos\left[N\theta - N\tan(\xi) \ln(R/R_0) \right]
\end{eqnarray}
where $f$ is the flutter factor, $N$ the number of sectors around the machine, $B_0$ the reference field at $R=R_0$ and $\xi$ is the spiral angle between the locus of the field maximum and the radius. Assuming small orbit scalloping around $R_0$ such that Eq. (\ref{Eq:F_phi}) holds, one obtains:
\begin{eqnarray}
q_y^{spi}(\theta) &=& - \left(1+\dot{\phi}\right) \dfrac{R}{F} \dfrac{\partial F}{\partial R} \approx - R \dfrac{\partial F}{\partial R} \label{Eq:qy_spir} \\
&=& -f N \tan(\xi) \sin\left[N\theta - N\tan(\xi) \ln(R/R_0) \right] \nonumber
\end{eqnarray}
At first sight, one might think that the average value of $q_y^{spi}$ would be zero. However, as will be established below, there is a net focusing effect due to the different path lengths of the particle in the focusing and defocusing fields. Under the assumption of small orbit scalloping, i.e. $\ln(R/R_0) \ll 1$, Eq. (\ref{Eq:qy_spir}) can be expanded:
\begin{eqnarray}
q_y^{spi}(\theta) \approx &-&f N \tan(\xi) \sin\left(N\theta\right) \nonumber \\
&+& f N^2 \tan^2(\xi) \ln(R/R_0) \cos\left(N\theta\right)
\label{Eq:qy_spi2}
\end{eqnarray}
Now, integrating Eq. (\ref{Eq:F_phi}) and choosing the reference such that $\phi(0)=0$ yields:
\begin{eqnarray}
\phi(\theta) \approx \int_{0}^{\theta} \left(F(R,u)-1\right) du \approx \dfrac{f}{N} \sin\left(N\theta\right)
\end{eqnarray}
Integrating again the expression of $\phi$ (Eq. (\ref{Eq:tan})), one can determine the azimuthal variation of the radius along the closed orbit:
\begin{eqnarray}
\ln\left(\dfrac{R(\theta)}{R_0} \right) = \ln\left(\dfrac{R(\theta=0)}{R_0} \right) - \int_0^{\theta} \phi(u) du \approx \dfrac{f}{N^2} \cos(N\theta) \nonumber
\end{eqnarray}
Injecting this into Eq. (\ref{Eq:qy_spi2}) yields:
\begin{eqnarray}
q_y^{spi}(\theta) \approx -f N \tan(\xi) \sin\left(N\theta\right) + f^2 \tan^2(\xi) \cos^2\left(N\theta\right) \nonumber
\end{eqnarray}
so that the first order contribution of the spiral focusing to the vertical tune is given by: 
\begin{eqnarray}
\mean{q_y^{spi}} = \dfrac{f^2}{2} \tan^2(\xi) = \mathcal{F}^2 \tan^2(\xi)
\end{eqnarray}
In addition, due to the Alternating Gradient focusing, a second contribution to the vertical focusing can be calculated:
\begin{eqnarray}
\mean{\widetilde{q_y^{spi}}^2} \approx \dfrac{f^2}{2} \tan^2(\xi) = \mathcal{F}^2 \tan^2(\xi)
\end{eqnarray} 
so that the overall contribution to the vertical tune is given by the often quoted formula \cite{craddock}:
\begin{eqnarray}
\nu_y^2 \approx -k + \mathcal{F}^2 \left[1+2\tan^2(\xi) \right]  
\end{eqnarray} \\ 
This formula is an approximation which becomes less and less accurate when the orbit scalloping becomes important and when the Alternating Gradient effect due to the average field index of the magnet becomes more and more important, i.e. for large $k$-values and for non-negligible values of $f$. \\

\section{Analytical expression of the magnetic field to account for radial defects}
\label{appendix:c}
In order to obtain the radial dependence of the field when the mean field index $k$ is R-dependent, let's assume that $k$ can be fitted with an $n$-order polynomial. Thus, $k$ writes in the following way:
\begin{eqnarray}
k(R)=\sum\limits_{i=0}^n a_i \left(\dfrac{R}{R_0}\right)^i   \label{Eq:14}
\end{eqnarray}
Then by equating Eq. (\ref{generalized_field_index}) and Eq. (\ref{Eq:14}), one obtains:
\begin{eqnarray}
\dfrac{dB}{B}=\sum\limits_{i=0}^n \dfrac{a_i}{R_0^i} R^{i-1}dR  \nonumber
            &=& a_0 \dfrac{dR}{R}+\sum\limits_{i=1}^n \dfrac{a_i}{R_0^i} R^{i-1}dR    \label{Eq:15}
\end{eqnarray} 
which gives after integration:
\begin{eqnarray}
\ln(\dfrac{B}{B_0})= a_0 \ln(\dfrac{R}{R_0})+\sum\limits_{i=1}^n \dfrac{a_i}{R_0^i} \int_{R_0}^{R} R^{i-1}dR  
\end{eqnarray}
\begin{eqnarray}
B(R)=B_0 \exp\left[a_0 \ln(\dfrac{R}{R_0})+ \sum\limits_{i=1}^n \dfrac{a_i}{R_0^i} \dfrac{(R^i-R_0^i)}{i}\right] \nonumber
\end{eqnarray}
so that the general form of the magnetic field becomes, $B(R,\theta)=B(R)F(\theta)$, i.e.,
\begin{eqnarray}
B(R,\theta)=B_0\left(\dfrac{R}{R_0}\right)^{a_0} \times \exp\left(\sum\limits_{i=1}^n a_i \dfrac{R^i-R_0^i}{i \times R_0^i}\right) \times F(\theta) \nonumber
\end{eqnarray}

\end{document}